\documentclass[twocolumn]{aastex61}
\pdfoutput=1
\date{\today}
\usepackage[utf8]{inputenc}
\usepackage{amsmath,amstext}
\usepackage[T1]{fontenc}
\usepackage{apjfonts} 
\usepackage{amssymb}
\usepackage[figure,figure*]{hypcap}
\newlength\figureheight
\newlength\figurewidth

\shorttitle{Dust in ESO\,323-G77}
\shortauthors{J.H. Leftley et al.}

\begin{document}
\author[0000-0001-6009-1803]{James H. Leftley}
\affiliation{European Southern Observatory, Alonso de Córdova 3107, Casilla 19001, Santiago, Chile}
\affiliation{Department of Physics \& Astronomy, University of Southampton, Southampton, SO17 1BJ, UK}
\author[0000-0001-8281-5059]{Konrad R. W. Tristram}
\affiliation{European Southern Observatory, Alonso de Córdova 3107, Casilla 19001, Santiago, Chile}

\author[0000-0002-6353-1111]{Sebastian F. H\"onig}
\affiliation{Department of Physics \& Astronomy, University of Southampton, Southampton, SO17 1BJ, UK}
\author[0000-0002-2216-3252]{Makoto Kishimoto}
\affiliation{Kyoto Sangyo University, Motoyama, Kamigamo, Kita-ku, Kyoto 603-8555, Japan}

\author[0000-0003-0220-2063]{Daniel Asmus}
\affiliation{Department of Physics \& Astronomy, University of Southampton, Southampton, SO17 1BJ, UK}
\affiliation{European Southern Observatory, Alonso de Córdova 3107, Casilla 19001, Santiago, Chile}
\author[0000-0003-3105-2615]{Poshak Gandhi}
\affiliation{Department of Physics \& Astronomy, University of Southampton, Southampton, SO17 1BJ, UK}

\title{New Evidence for the dusty wind model: \\ Polar dust and a hot core in the type-1 Seyfert ESO\,323-G77\footnote{Based on European Southern Observatory (ESO) observing programmes 083.B-0452, 084.B-0366, 087.B0401, 092.B-0718, 095.B-0376, and 290.B-5113.}} 

\begin{abstract}

Infrared interferometry of Seyfert galaxies has revealed that their warm ($300-400\,$K) dust emission originates primarily from polar regions instead of from an equatorial dust torus as predicted by the classic AGN unification scheme. We present new data for the type 1.2 object ESO\,323-G77 obtained with the MID-infrared interferometric Instrument (MIDI) and a new detailed morphological study of its warm dust. The partially resolved emission on scales between 5 and 50\,mas ($1.6-16$\,pc) is decomposed into a resolved and an unresolved source. Approximately $65\%$ of the correlated flux between $8$ and $13\,\mu\mathrm{m}$ is unresolved at all available baseline lengths. The remaining $35\%$ is partially resolved and shows angular structure. From geometric modelling, we find that the emission is elongated along a position angle of $155^\circ\pm14^\circ$ with an axis ratio (major/minor) of $2.9\pm0.3$. Because the system axis is oriented in the position angle $174^\circ\pm2^\circ$, we conclude that the dust emission of this object is also polar extended. A \textit{CAT3D-WIND} radiative transfer model of a dusty disk and a dusty wind with a half opening angle of $30^\circ$ can reproduce both the interferometric data and the SED, while a classical torus model is unable to fit the interferometric data. We interpret this as further evidence that a polar dust component is required even for low-inclination type 1 sources.
\end{abstract}

\section{Introduction}

The distribution of hot, $\approx1000$\,K, and warm, $300-400$\,K, dust in Active Galactic Nuclei (AGN) has come under scrutiny in recent years with improvements in long baseline infrared (IR) interferometry. Thermal emission from warm dust dominates the mid-IR emission in the central region of AGN peaking in the $8-13\,\mu\mathrm{m}$ range. In the classical model of AGN, the bulk of the warm dust is housed in the obscuring dust torus around the equatorial region of the AGN. This torus is responsible for the obscuration required by the unification scheme \citep{antonucci_unified_1993}. With high angular resolution mid-IR interferometers, such as the MID-infrared interferometric Instrument, or MIDI \citep{leinert_midi_2003}, and the upcoming Multi-AperTure mid-Infrared SpectroScopic Experiment, or MATISSE \citep{lopez_matisse:_2008}, we can directly observe AGN on the scale of the dust torus, $1-10\,$pc, for the first time \citep[e.g][]{tristram_mid-infrared_2007,kishimoto_mapping_2011,honig_parsec-scale_2012,honig_dust_2013,burtscher_diversity_2013,tristram_dusty_2014,honig_dust-parallax_2014,lopez-gonzaga_mid-infrared_2016}. In type 1 AGN, where the Broad Line Region (BLR) is directly visible, we would expect very little angular dependence on the size of the emission source -- presumably the dust torus -- due to projection effects. On the other hand, as we move toward type 2 AGN, we would expect to see an equatorial extension from where the bulk of the dust is located.

However, it turns out that the warm dust emission instead extends in the polar direction. This finding cannot be explained by the classical model of AGN; yet, the unification scheme requires a source of obscuration. This led to the proposal of the disk+wind model \citep{honig_dust_2013,honig_dusty_2017}. In this model, the warm torus is replaced by a wind driven polar dust structure and a dust disk.

Here, we aim at constraining the distribution of dust in the centre of an unobscured AGN, the Seyfert 1.2 galaxy ESO\,323-G77, using mid-infrared interferometry. ESO\,323-G77 is located at a distance of $\sim60\,$Mpc ($0.311\,\mathrm{pc}\,\mathrm{mas}^{-1}$) and has been studied with interferometry before. In \citet{kishimoto_mapping_2011}, four \textit{uv} points were fitted with a power-law brightness distribution which revealed that the size of the emission area decreases toward shorter wavelengths. \citet{burtscher_diversity_2013} found a significantly smaller size of the emission than \citet{kishimoto_mapping_2011} using a Gaussian model with the same four observations and one more that was deemed not usable in \citet{kishimoto_mapping_2011}. \citet{lopez-gonzaga_mid-infrared_2016} performed a more detailed statistical analysis of these data using an elongated Gaussian plus point source model. While formally finding a Full Width at Half Maximum (FWHM) of the minor axis of 17\,mas (5.3\,pc) and an axis ratio of $1.4$, the \textit{uv} plane was determined to be insufficient to robustly constrain the elongation and its position angle (PA). \citet{asmus_subarcsecond_2016} reports an extension along the putative torus mid-plane at subarcsecond spatial resolution based on single-telescope mid-IR images, but also point out that this is likely to be an observational artefact. Such an extension would be on larger scales than that of the putative dust torus.

In this paper, we present 10 successful mid-IR interferometric \textit{uv} data points in order to pin down the radial and angular-dependent size of the warm dust in ESO\,323-G77. In Section \ref{Observations}, we describe the observations and data reduction, including new measurements with the VLT-mounted Imager and Spectrometer for the mid-IR (VISIR; \citealt{lagage_successful_2004, kaufl_return_2015, kerber_visir_2016}). In Section \ref{Results}, we present our findings from the reduced VISIR data as well as an initial interpretation of the MIDI data. In Section \ref{Data Analysis}, we model the MIDI data with, firstly, a geometric model and, secondly, the radiative transfer model \textit{CAT3D-WIND}. In Section \ref{Discussion}, we discuss our results in the context of other measurements. Finally, a summary is given in Section \ref{Summary}.

\section{Observations}\label{Observations}

\subsection{Interferometric Data}

The interferometric observations for ESO\,323-G77 used MIDI \citep{leinert_midi_2003} at the Very Large Telescope Interferometer (VLTI) facility. MIDI is a two beam mid-IR interferometer that combines beams from pairs of telescopes. It covers the wavelength range of 8$-$13 microns, \textit{N}-band, accessible from the ground. 

\subsubsection{Interferometric Observations} \label{S I O}

For all MIDI observations, both those from the previous works and the additional new data, only 8.2\,m unit telescope (UT) pairs were used due to the relative faintness of ESO\,323-G77. The observations were carried out between 2009 and 2014.

\begin{table*}
\caption{Table of MIDI observations of ESO\,323-G77.}\label{tab Observation}
\begin{tabular*}{\textwidth}{l @{\extracolsep{\fill}} c c c c c c c c c}\hline
Date&UT&Stations&Proj BL (m)&PA (deg)&Calibrator&Comment&Programme&Note&No. Files\\ \hline \hline
2009 May 08&01:04:03&UT3/UT4&56.8&94.4&HD112213&&083.B-0452&a&2\\
2009 May 10&03:12:36&UT1/UT2&54.1&30.5&HD112213&&083.B-0452&a&2\\
2009 May 10&04:15:43&UT1/UT2&51.8&37.1&HD112213&&083.B-0452&a&2\\
2009 May 11&04:07:36&UT1/UT4&118.4&75.0&HD112213&&083.B-0452&a&1\\
2009 May 12&00:10:57&UT1/UT4&126.9&36.9&HD112213&&083.B-0452&a&2\\
2010 Mar 01&03:15:11&UT1/UT2&56.3&353.2&HD112213&&084.B-0366 &b&2\\
2010 Mar 02&04:43:06&UT1/UT3&102.4&11.6&HD112213&&084.B-0366 &b&2\\
2010 Mar 02&06:00:16&UT1/UT3&101.8&23.5&HD112213&&084.B-0366 &b&2\\
2010 Mar 02&07:50:55&UT1/UT3&97.3&38.5&HD112213&Poor fringe tracking&084.B-0366&b&1\\
2011 Apr 15&05:02:59&UT2/UT3&44.5&47.7&HD100407&&087.B-0401&b&2\\
2014 Mar 16&06:23:41&UT1/UT3&99.5&34.0&HD112213&&092.B-0718&b&2\\ \hline
\end{tabular*}
\tablecomments{(a) Data previously used in \citet{kishimoto_mapping_2011}, \citet{burtscher_diversity_2013}, and \citet{lopez-gonzaga_mid-infrared_2016}; (b) Unpublished data.}
\end{table*}

In total, 11 fringe tracks and complimentary single-dish spectra were observed. All observations are listed in Table \ref{tab Observation}. HD112213 was used as the calibrator star except on 2011 April 15, where HD100407 was observed instead. We supplemented all our single-dish spectra with the high S/N VISIR spectrum from \citet{honig_dusty_2010-1}, observed on 2009 May 04, which provides a better reference to determine visibilities and therefore sizes (see Section ~\ref{sec:visir}).

\subsubsection{Interferometric Data Reduction}

The MIDI observations were reduced using the coherent visibility estimation as implemented in \textsc{ews} \citep{jaffe_coherent_2004}; further analysis of the data was performed using \textsc{python 2.7} with the \textsc{astropy}, \textsc{emcee}, and \textsc{oifits} libraries \citep{astropy_collaboration_astropy:_2013,foreman-mackey_emcee:_2013}. All observations from previous works were re-reduced in order to obtain a homogeneous data set.

For the data reduction, we followed \citet{burtscher_observing_2012}. The total flux, $F_\mathrm{tot}(\lambda)$, and the correlated flux, $F_\mathrm{corr}(\lambda)$ were extracted using an optimised mask for each observing epoch. The widths of these masks increase from $5.0\pm0.35\,\mathrm{pixels}$ ($0.42"$) at $8\,\mathrm{\mu m}$ to $7.7\pm0.8\,\mathrm{pixels}$ ($0.65"$) at $13\,\mathrm{\mu m}$.

\begin{figure}
\includegraphics[width=0.5\textwidth, trim={1cm 0 0 0}, clip]{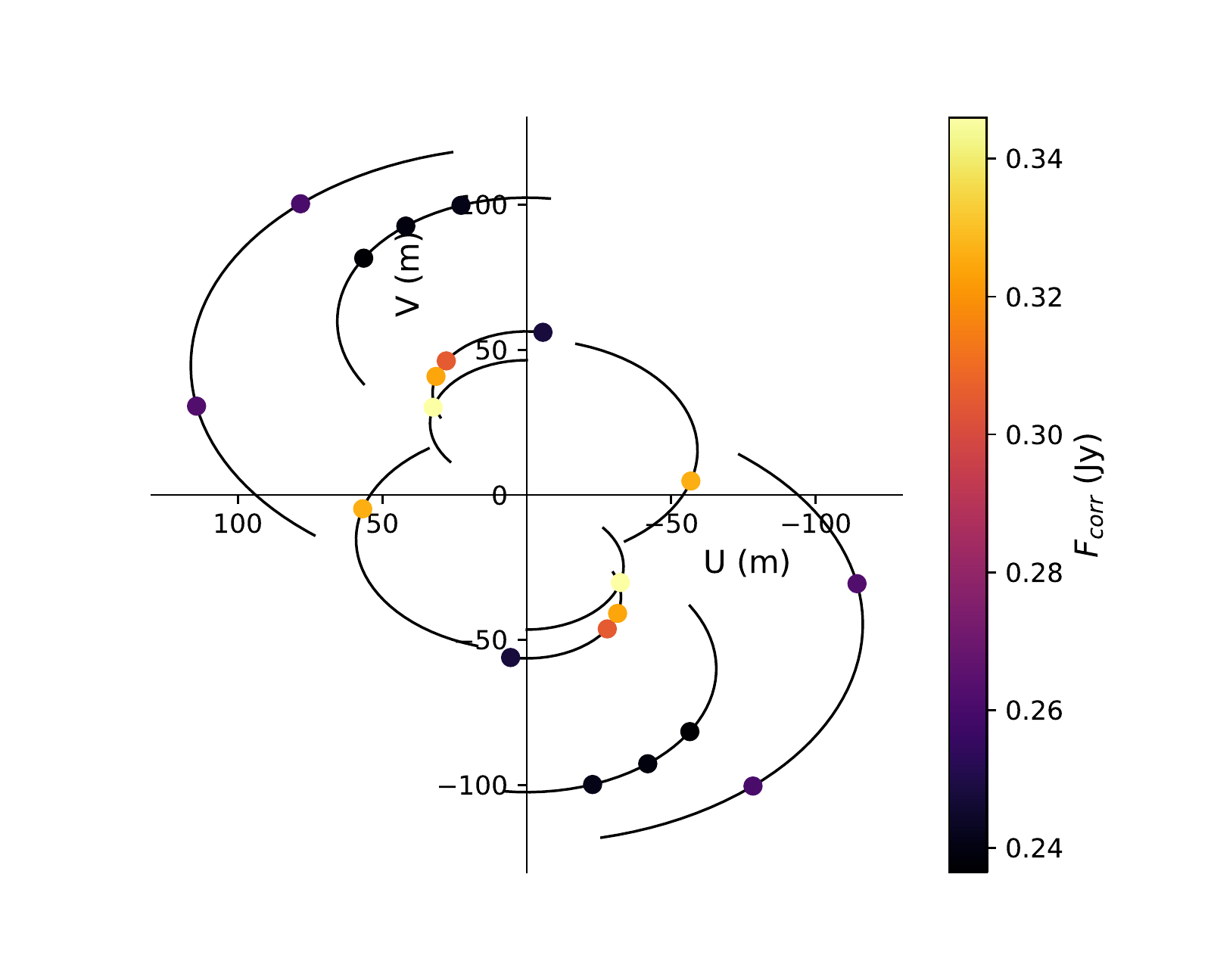}
\caption{The \textit{uv} coverage for ESO\,323-G77. The individual points are colour coded according to their correlated flux in the $11.8\,\mu$m bin, $F_\mathrm{corr} (11.8\,\mathrm{\mu m})$.}
\label{uv plane}
\end{figure}

There were 25 attempted fringe tracks on ESO\,323-G77 with MIDI in total. One of these, however, had an insufficient signal for fringe tracking due to high wind (see Table \ref{tab Observation}), three, on 2009 March 09, have insufficient signal to noise, because the fringes could not be tracked properly; and two on 2014 March 16 are non-detections due to extreme atmospheric dust. The remaining 19 observations were sorted into 30 minute groups, combined, and reduced in accordance with \citet{lopez-gonzaga_mid-infrared_2016}. We hence obtain 10 independent measurements of $F_\mathrm{corr}(\lambda)$, as listed in Table \ref{tab Observation} and shown in Figure \ref{uv plane}.

By default, the calibrator observed closest in time to the science target was used for the calibration. We accounted for the error of the transfer function from intra-night fluctuations by creating a structure function (SF) from every calibrator taken throughout the same night as our science target. The value of the SF for the time difference between the science object and its calibrator was added in quadrature to the statistical error provided by \textsc{ews}. For each night, the SF showed a clear structure consistent with non-random variation, most likely due to variations of the observing conditions.

\begin{figure}[t]
\includegraphics[width=0.5\textwidth]{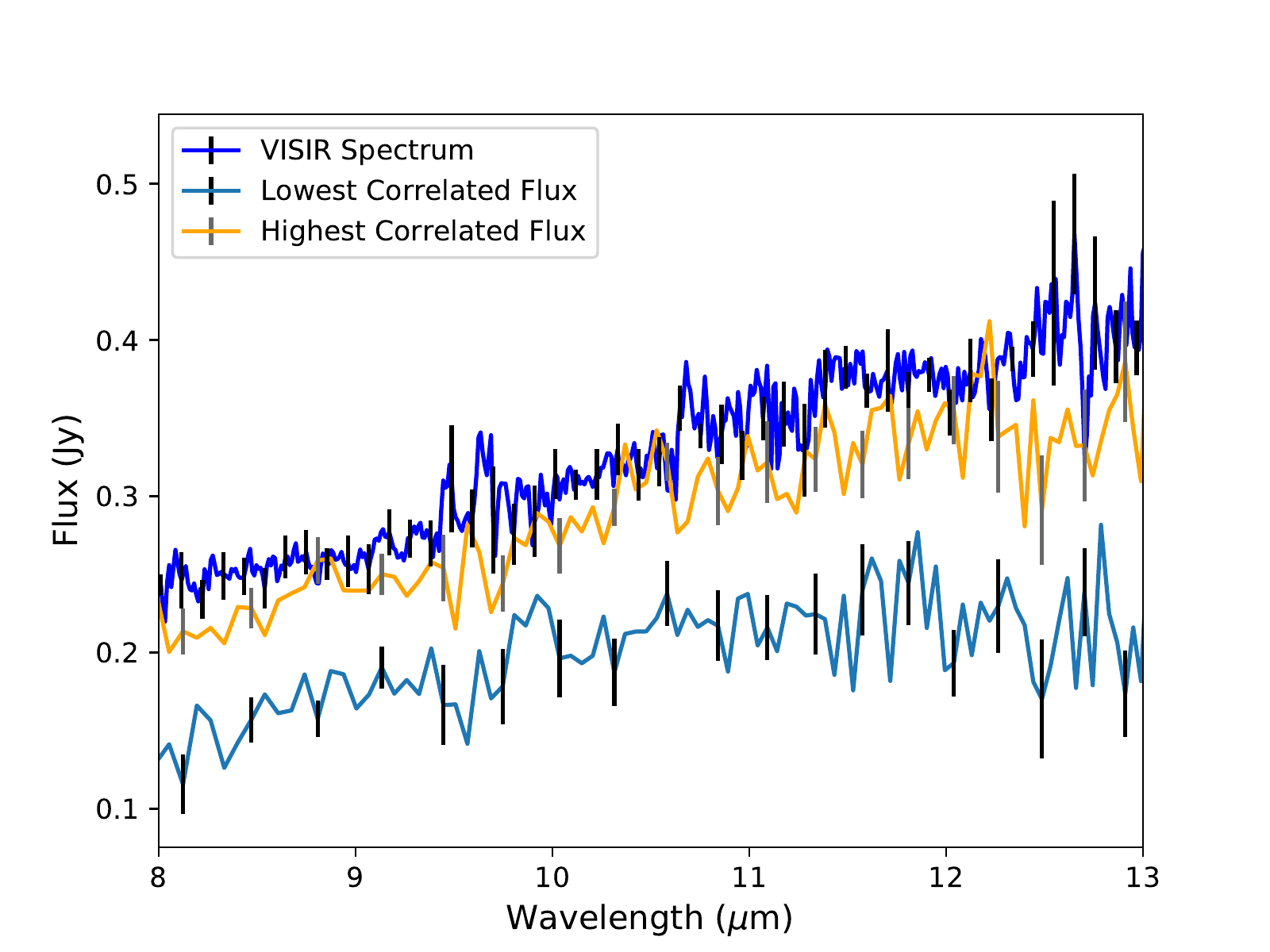}
\caption{The highest and lowest correlated fluxes measured by MIDI. The VISIR spectrum is also shown for comparison.}
\label{corr flux comp}
\end{figure}

The uncertainties of the MIDI single-dish spectra were too large to make them usable. Instead, we used the total flux spectrum from \citet{honig_dusty_2010-1} obtained with VISIR, $F_\mathrm{VISIR}(\lambda)$, to calculate the visibility using: $V(\lambda)=\frac{F_\mathrm{corr}(\lambda)}{F_\mathrm{VISIR}(\lambda)}$. Because the VISIR spectrum was extracted by fitting a Gaussian to the PSF with $\mathrm{FWHM} = 0.4"$ at $12\,\mu$m, the extraction window is similar to the mask used for MIDI, and hence aperture effects should be minimal, considering that the source is essentially unresolved in the mid-IR (see Section \ref{morphology}). The VISIR spectrophotometry, as well as the highest and lowest $F_\mathrm{corr}(\lambda)$ values, are plotted in Figure \ref{corr flux comp}. We compared our re-reduced data to that of \citet{burtscher_diversity_2013} and found that the resulting $F_\mathrm{corr}(\lambda)$ values are in agreement within errors.

\subsection{VISIR Data}
\label{sec:visir}

ESO\,323-G77 was observed with the upgraded VISIR as part of programme 095.B-0376 to study mid-IR flux changes in the last 6 years since its previous spectrophotometry with VISIR \citep{honig_dusty_2010-1, asmus_subarcsecond_2014}. 
The observations were carried out in standard imaging mode in the PAH1 ($8.6\,\mu\mathrm{m}$) and PAH2\_2 ($11.88\,\mu\mathrm{m}$) filters with perpendicular nodding and a chop/nod throw of $8"$. 
In total, four epochs were obtained with each filter in service mode between 2015 August and 2016 February with on-source exposure times of 7 and 10 minutes, respectively (see Table \ref{tab VISIR Observation}). For flux calibration and point spread function (PSF) reference purposes, the science observations were preceded and followed by a mid-IR standard star from the catalogue from \citet{cohen_spectral_1999}.

The data reduction was performed with a custom made \textsc{python} pipeline, and flux measurements were obtained using the custom developed \textsc{idl} software package, \textsc{mirphot} \citep{asmus_subarcsecond_2014}. The results are displayed in Table \ref{tab VISIR Observation}.

\subsection{Finding the System Axis}

The system axis in this paper is the polar axis, i.e.\ the axis that is perpendicular to the accretion disk, implied by the optical polarisation and the Narrow Line Region (NLR). \citet{mulchaey_emission-line_1996} presented spatially extended [\ion{O}{3}] emission from the NLR of ESO\,323-G77. However, in type 1 Seyfert galaxies the central, unresolved, [\ion{O}{3}] emission often drowns out the more extended, lower surface-brightness, NLR emission resulting in poorly constrained extensions. Thus, the authors only note a North$-$South extension. New data from the S7 survey reveals the [\ion{O}{3}] emission in lower spatial resolution but with higher S/N \citep{thomas_probing_2017}. The pre-reduced data products are available at \url{https://datacentral.aao.gov.au/asvo/surveys/s7/}. We used the line fitting program \textsc{kubeviz} \citep{fossati_muse_2016} to perform a single Gaussian and continuum fit to the 500.7\,nm and 495.9\,nm [\ion{O}{3}] lines. The 500.7\,nm component is shown in Figure \ref{O3} with the system axis overplotted. Polarised emission in AGN arises from either an equatorial scatterer on the scale of the BLR inside the putative torus, or the polar region co-spatial with the innermost NLR. \citet{schmid_spectropolarimetry_2003} measured a polarisation angle of the optical emission of $84^\circ\pm2^\circ$, this is reaffirmed with near-IR polarisation measurements from \citet{batcheldor_nicmos_2011}. \citet{smith_seyferts_2004} reports that the polarisation in ESO\,323-G77 is dominated by scattering in the polar region. Therefore, the system axis is expected at an angle of approximately $174^\circ\pm2^\circ$ in agreement with the polar axis inferred from the NLR.

\begin{table}
\caption{VISIR photometry of ESO\,323-G77.}\label{tab VISIR Observation}
\begin{tabular*}{0.47\textwidth}{l @{\extracolsep{\fill}} c c c c}\hline
Date&Filter&$\lambda(\mu$m)&Flux (mJy)&Error (mJy)\\ \hline \hline

2009 May 10&ARIII&8.99&300.9&31.6\\
2010 Mar 10&PAH1&8.59&275.2&31.6\\
2015 Aug 03$^a$&PAH1&8.59&315&54.2\\
2016 Jan 15&PAH1&8.59&369.1&38.1\\
2016 Feb 16&PAH1&8.59&340.5&34.5\\
2016 Feb 28&PAH1&8.59&359.9&36.6\\

2009 May 10&PAH$2\_2$&11.88&383.4&40.7\\
2010 Mar 10&PAH$2\_2$&11.88&378.7&40.3\\
2015 Aug 03$^a$&PAH$2\_2$&11.88&695&374\\
2016 Jan 15&PAH$2\_2$&11.88&440.0&44.7\\
2016 Feb 16&PAH$2\_2$&11.88&473.4&54.4\\
2016 Feb 28&PAH$2\_2$&11.88&449.2&45.5\\
\hline
\end{tabular*}
\tablecomments{$^a$ Large uncertainties due to strong background variations from clouds.}
\end{table}

\section{Results}\label{Results}

\subsection{Morphology and Variability from Single-dish Observations}\label{morphology}

The previous subarcsecond resolution mid-IR images of ESO\,323-G77 from \citet{asmus_subarcsecond_2016} indicate that the nucleus may be marginally extended in comparison to the PSF as measured from consecutively observed standard stars with major and minor axis FWHMs of $0.40"$ and $0.38"$ at $12\,\mu\mathrm{m}$. The PA of the major axis is, on average, $95^\circ$ but with a significant scatter of $27^\circ$, so that any elongation remains uncertain. Our new data show the same morphology at $12 \,\mu\mathrm{m}$, while being consistent with an unresolved nucleus at $8.6\,\mu\mathrm{m}$. The variations in the PSF as measured from the standard stars before and after are too large to draw any firmer conclusion on the nuclear mid-IR morphology. Even the average combined PAH1 and PAH2\_2 images, with total exposure times of 30 and 40\,minutes, respectively, do not reveal any further, faint structures.

Furthermore, the VISIR imaging from \cite{asmus_subarcsecond_2014}, obtained contemporaneous with the 2009 and 2010 MIDI observations, does not reveal any significant flux changes. From the new measurements obtained between 2015 April and 2016 February, we derive average flux densities of the total nuclear emission of $346 \pm 21\,$mJy in PAH1 and $514 \pm 96\,$mJy in PAH2\_2 for all epochs combined, without any sign of evolution during the seven months the observations were taken. However, these values are 20\% higher than the values measured in 2009/2010. Therefore, it is likely that the intrinsic flux of the AGN increased over the last $5-6$ years. Because we use a common VISIR spectrum to find $V(\lambda)$, this could cause $V(\lambda)$ to be overestimated in observations after 2010, especially effecting the 2014 observation. We therefore compared the 2014 MIDI observation to the two close UT1/UT3 observations from 2010. $F_\mathrm{corr}(\lambda)$ is lower for the 2014 observation than the two 2010 observations, which would not be expected if it were effected by an \emph{increase} in flux. Furthermore, we compared observations of the same PA with different baseline lengths. Again, the 2014 measurement lies below a Gaussian interpolation of the measurements from 2010. We therefore conclude that the increase in $F_\mathrm{tot}(\lambda)$ did not lead to a measurable increase of $F_\mathrm{corr}(\lambda)$ and hence does not significantly impact our results.

\subsection{Interferometric Observations}

\textsc{ews} provides wavelength differential phases, the absolute phase information being destroyed by the atmosphere. However, the differential phase spectrum for every observation is flat within $1\sigma$ errors. This tells us that there is no image centre shift in the wavelength range observed. Therefore, we did not include phase information in our fitting beyond making the model's centre invariant in location and the model centrosymmetric.

\subsubsection{Dust Morphology from Interferometric Observations}
\begin{figure}[tbp]
\includegraphics[width=0.47\textwidth, trim={0.7cm 0 0 0},clip]{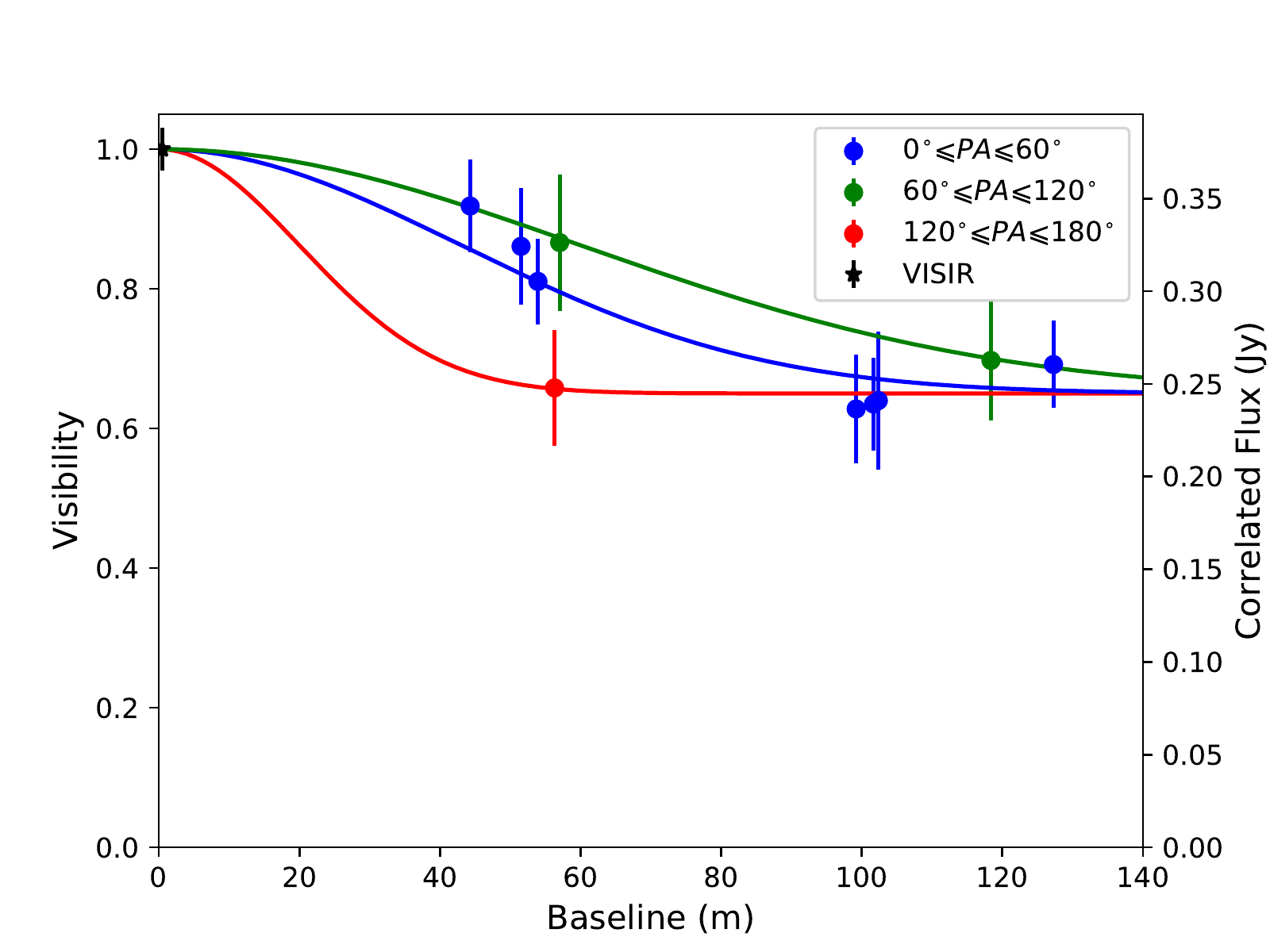}
\caption{$V(11.8\,\mu\mathrm{m})$ observations binned into three groups based on PA and fitted with a Gaussian and a constant. The singular red point was given the same constant as the other two fits.}
\label{fig BL ext}
\end{figure}

As an initial check for structure, we plot $V(\lambda)$ against BL in Figure \ref{fig BL ext}. $V(11.8\,\mu\mathrm{m})$ never falls below 0.6, which makes the source at least 60\% unresolved. An interesting feature of this plot is an apparent levelling off of $V(\lambda)$ for $\mathrm{BL}>80\,\mathrm{m}$. This could be interpreted as the source having two distinct components, one of which is partially resolved and the second unresolved and responsible for more than 60\% of the flux. This unresolved component must be on a scale smaller than \textasciitilde$5\,$mas.

To check for any angular dependence, we binned the observations, by PA, into three bins of $60^{\circ}$ (Figure \ref{fig BL ext}). When considering each bin separately, we see a clear decrease with BL in the green and blue bins. The remaining red bin contains only one observation; therefore, we cannot see the decrease for this bin directly. However, we know it has a visibility of 1 at a BL of $0-8.2\,$m from the VISIR single dish imaging and we assume that the unresolved component does not show any noticeable angular dependence. We assigned each bin a Gaussian and a constant to show the baseline dependant change. All the constants were set to be the same at 0.65. In Figure \ref{fig BL ext} we can see that the source is more resolved in the $120^{\circ}-180^{\circ}$ bin and less resolved in the $60^{\circ}-120^{\circ}$ bin.

\begin{figure}[tbp]
\includegraphics[width=0.5\textwidth, trim={2cm 0cm 1cm 0cm}, clip]{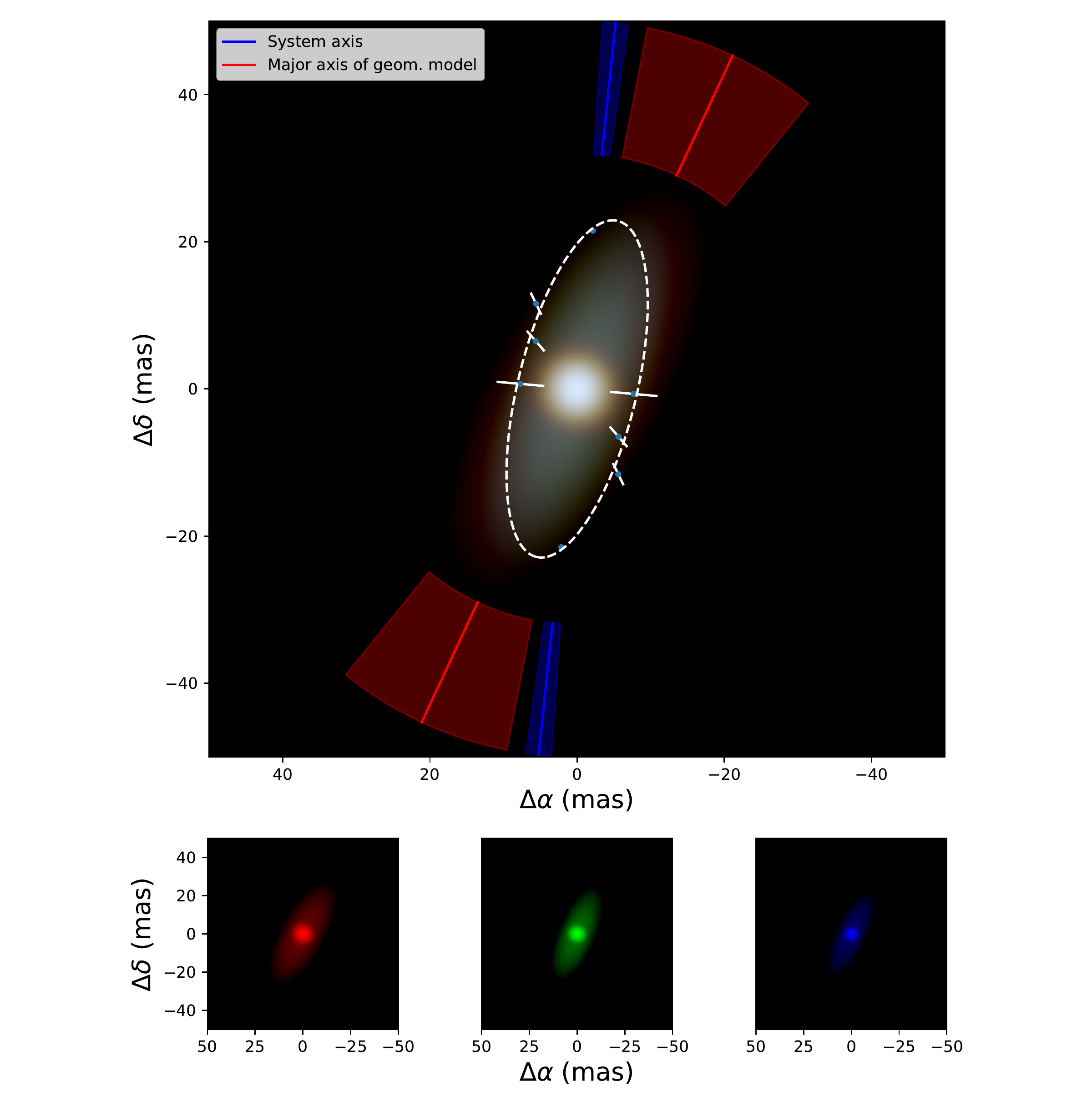}
\caption{False colour image of the geometric model. The blue, green, and red channels are the average geometric model fits for $8.6\,\mu\mathrm{m}$ and $9\,\mu\mathrm{m}$, $10.2\,\mu\mathrm{m}$, and $10.6\,\mu\mathrm{m}$, and $11.8\,\mu\mathrm{m}$ and $12.2\,\mu\mathrm{m}$ respectively. Each channel is plotted individually below the main panel. The models fluxes are plotted with logarithmic colour scaling. Overplotted is the 66\% light radius at $11.8\,\mu\mathrm{m}$ for four different position angles, fitted with an ellipse (dashed line). Also indicated is the mean position angle of the major axis of the geometrical model and its uncertainty in red (c.f. Figure \ref{test results} and \ref{PA-Lam}) and the system axis in blue.}
\label{66lightrad}
\end{figure}

We therefore performed a clearer, model independent, initial test for angle dependant structure by calculating the 66\% light radius ($r_{0.66}$) for the $11.8\,\mu$m data shown in Figure \ref{fig BL ext}. We chose 66\% of the flux to match the flux value of the singular red point. We separated the observations into four bins at $26^\circ$, $41^\circ$, $85^\circ$, and $174^\circ$ with 4, 3, 2, and 1 observations in each bin respectively. We then fit a straight line to each bin, except for the $174^\circ$ bin, which corresponds to the lone red point in Figure \ref{fig BL ext}, and interpolated each of the other bins to the visibility value of the red point, $V=0.66$. This provides us with the BL length at which each bin has a visibility value of 0.66. We convert this to RA and DEC, using $r_{0.66}=\frac{BL}{2\lambda}$, and derive errors for the fitted bins using a Monte Carlo method. The result of this is the 66\% light radius, which is the angular size at which 66\% of the flux is contained. To this we fit an ellipse. The result is plotted in Figure \ref{66lightrad} with a PA of the major axis of $166^\circ$ and an axis ratio of 3.1.

\section{Modelling}\label{Data Analysis}
\subsection{Geometric Model}
\label{Geomodel}

Due to the low \textit{uv} coverage and the lack of absolute phase information, it is impossible to reconstruct an image from our data; instead, we have to fit a model. To obtain a first understanding for the geometry of the warm dust in ESO\,323-G77, we fit a two component model to the visibility data, similar to the one described in \citet{lopez-gonzaga_mid-infrared_2016}. In our model, the unresolved component is modelled by a 2D radially symmetric Gaussian with an FWHM of the dust sublimation radius of ESO\,323-G77 given in \citet{kishimoto_mapping_2011}. The second Gaussian is an elongated 2D Gaussian. The model was analytically converted to Fourier space to produce the final model for fitting. The analytical description of the visibility is given by:
\begin{center} 
\[V(u,v,\lambda) = (1-p_f)F_1(u,v,\lambda)+p_fF_2(u,v,\lambda),\]
where
\[F_1(u,v,\lambda)=\exp\left[-\left(\frac{ C \Theta_y }{\lambda }\right)^2 (\frac{v'^2}{\epsilon^2}+u'^2)\right],\]
\[F_2(u,v,\lambda)=\exp\left[-\left(\frac{ C \Theta_s }{\lambda }\right)^2 (v^2+u^2)\right],\]
and
\[C=\frac{\pi^2}{1.296\cdot10^9\cdot\,\sqrt[]{\ln2}},\]
\[v'= u \cos\theta+v \sin\theta,\]
\[u'= u \sin\theta-v \cos\theta,\]
\end{center}
where $\Theta_y$ is the FWHM value of the major axis of the elongated Gaussian in mas, $\epsilon$ is the major to minor axis ratio of the elongated Gaussian, $\Theta_s$ is the sublimation radius in mas, $\lambda$ is the wavelength, and $p_f$ is the unresolved fraction. The constant $C$ comes from the conversion of degrees to mas and the conversion from $\sigma$ to FWHM with an extra factor of $\pi$ originating from the Fourier transform.

The $V(\lambda)$ measurements were separated into bins by wavelength with a width of $0.4 \,\mu\mathrm{m}$ and fitted independently of each other. This had the advantage of being able to test the models dependence on $\lambda$.

To reduce the bias effect from the under-sampled directions in the \textit{uv} plane, we used MCMC Bayesian model fitting with a flat prior as implemented by \textsc{emcee} \citep{foreman-mackey_emcee:_2013}. This fitting method had the advantage of providing probability distributions for all fitted parameters. Using MCMC Bayesian fitting does not necessarily make our fit reliable, the fitting had to be adequately tested.

\subsection{Influence of the \textit{uv} Plane} \label{influv}
\begin{figure}
\includegraphics[width=0.5\textwidth]{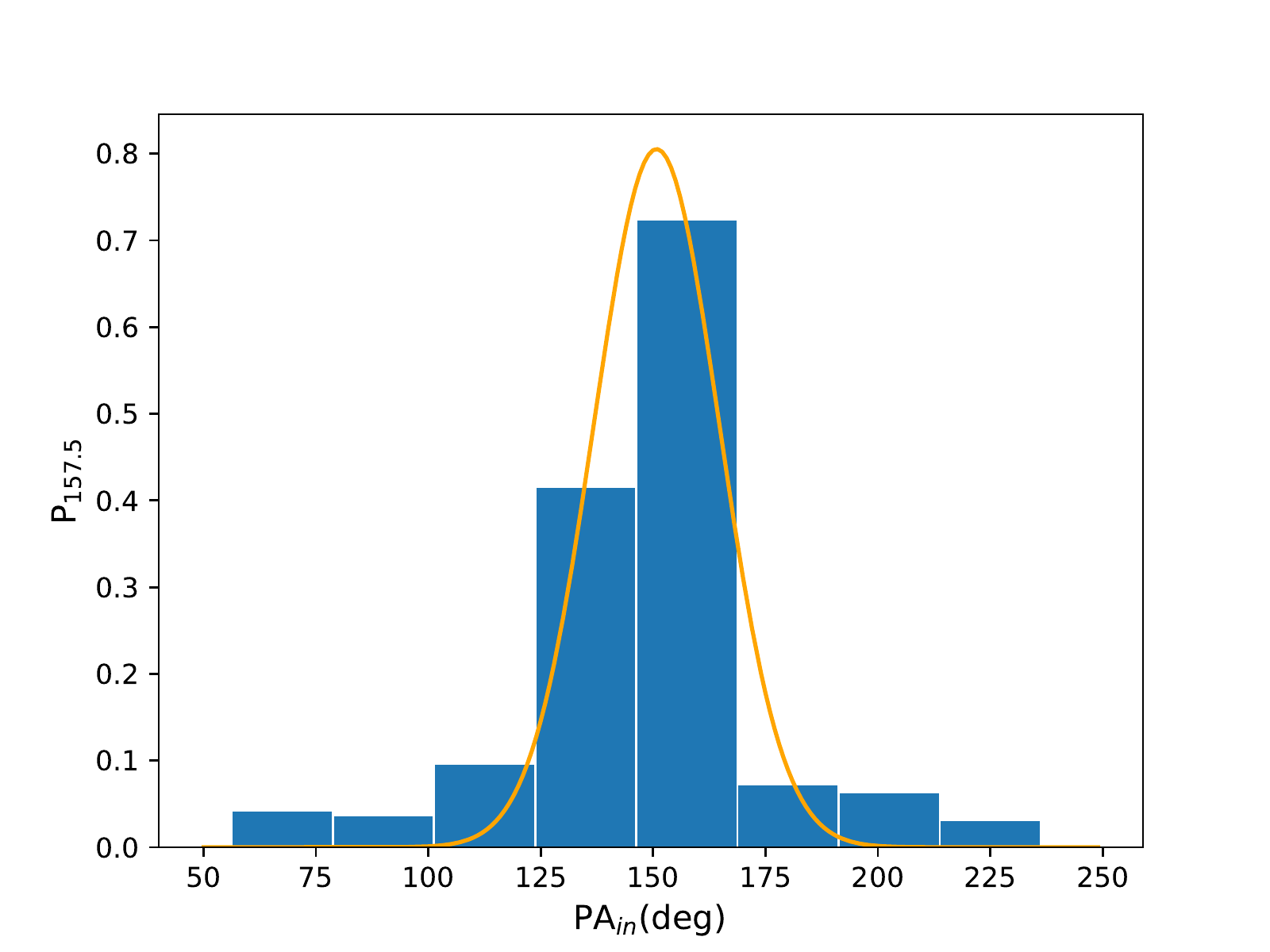}

\label{test results}
\caption{Frequency with which an object with position angle, PA, of PA$_{\mathrm{in}}$ is modelled with a PA between $146^\circ$ and $169^\circ$. This gives the uncertainty of the fitted PA of ESO\,323-G77 at the 68\% and 95\% confidence intervals as $14^\circ$ and $29^\circ$ respectively. This does not include objects with an axis ratio, $\epsilon$, of 1.}
\end{figure}

To ensure any results produced were uninfluenced by the positions of the \textit{uv} points, we input a wide range of test data. 
We employed the same method of mock data creation as that in \citet{lopez-gonzaga_mid-infrared_2016}. The test data was produced by using the same model as was fitted to the data. We used the \textit{uv} position values from the real data and assigned each a mock visibility from the mock model. Each \textit{uv} point was then randomly offset by Gaussian noise derived from the real error of the original it represented. We then fitted this set of mock data and compared the resulting parameters to the input parameters to test the reliability of the recovered PA, not the viability of the model itself. To make sure there were no degeneracies in the fitting due to poor \textit{uv} coverage, all parameter combinations were tested, removing any degenerate pairings. Taking into account the fit result from \citet{burtscher_diversity_2013} and \citet{lopez-gonzaga_mid-infrared_2016}, we chose the following parameter space: \[0.0\leqslant p_f \leqslant 0.9 \: \mathrm{in\ steps\ of} \: 0.1 \:\] \[1.0\leqslant \epsilon \leqslant 4.0 \: \mathrm{in\ steps\ of} \: 1.0 \: \] \[11.7\leqslant \Theta_y \leqslant 49.0 \: \mathrm{\ in\ steps\ of\ } \: 9.3 \: \] \[0\fdg0\leqslant PA \leqslant 157\fdg5 \: \mathrm{\ in\ steps\ of\ } \: 22\fdg5 \: \]
This was repeated five times, giving 6300 results; the result for PA is plotted in Figure \ref{test results}. From the test, we derive an uncertainty for our recovered PA of $14^{\circ}$ for $68\%$ confidence and $29^{\circ}$ for $95\%$ confidence. We conclude that with our new measurements since 2010 the \textit{uv} coverage is now sufficient to constrain any elongation.

\subsection{Geometric Modelling Results}
The results of the MCMC Bayesian fitting of the interferometric data at each wavelength can be found in Table \ref{Tresult}. The detailed fit results can be found in Figures \ref{8.2}$-$\ref{13.0} in the Appendix. We find that the fits at longer wavelength, $\lambda >10 \,\mu\mathrm{m}$, are well constrained. At shorter wavelengths, we find that the UT3/UT4 observation dominates, making the fit less constrained.

\begin{figure}[tbp!]
\begin{center}
\includegraphics[width=0.5\textwidth]{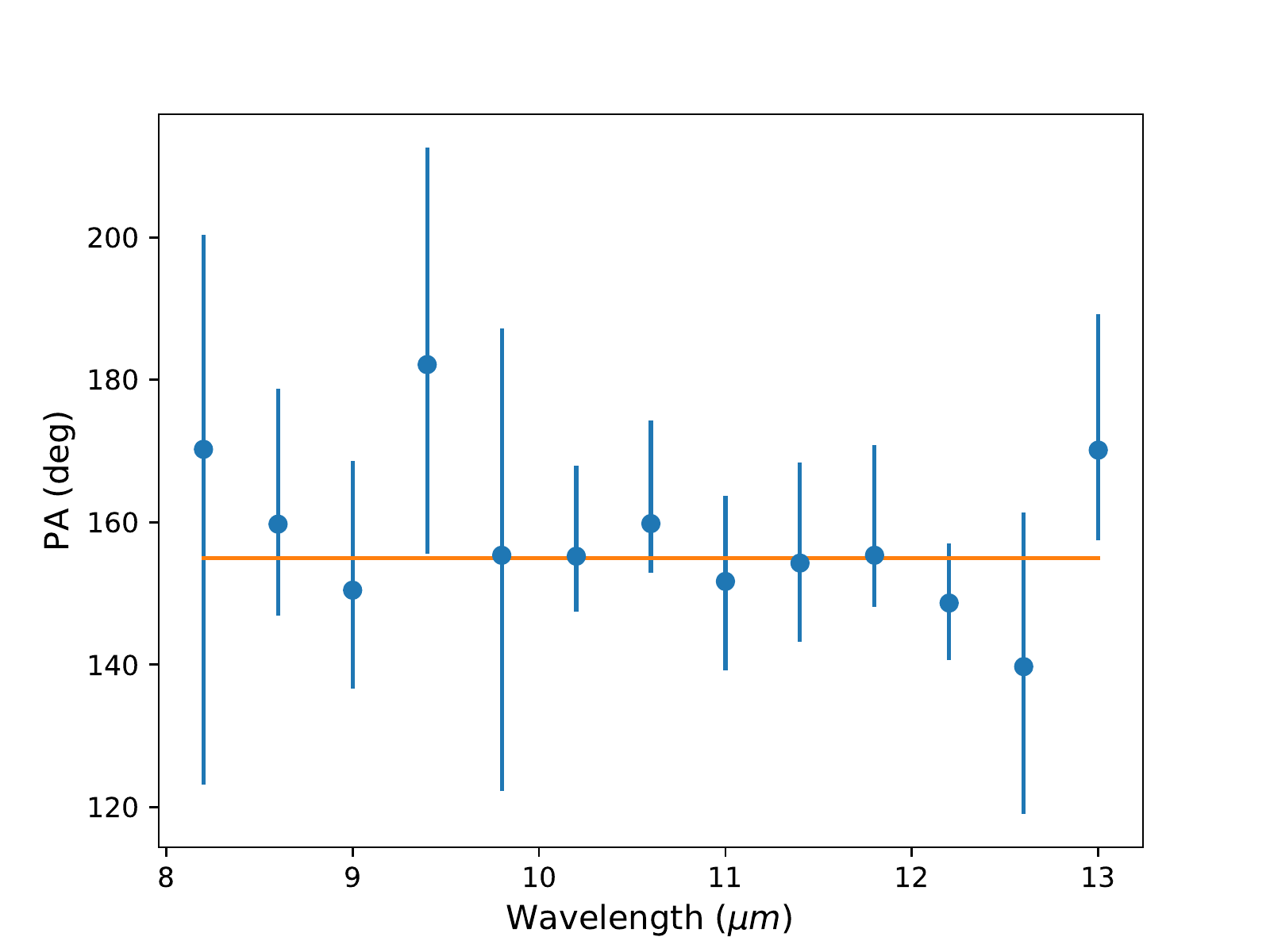}
\caption{Position angle, PA, of the major axis of the geometric model for each wavelength bin. The orange line is the weighted average value.}
\label{PA-Lam}
\end{center}
\end{figure}

\begin{figure}[tbp!]
\begin{center}
\includegraphics[width=0.5\textwidth]{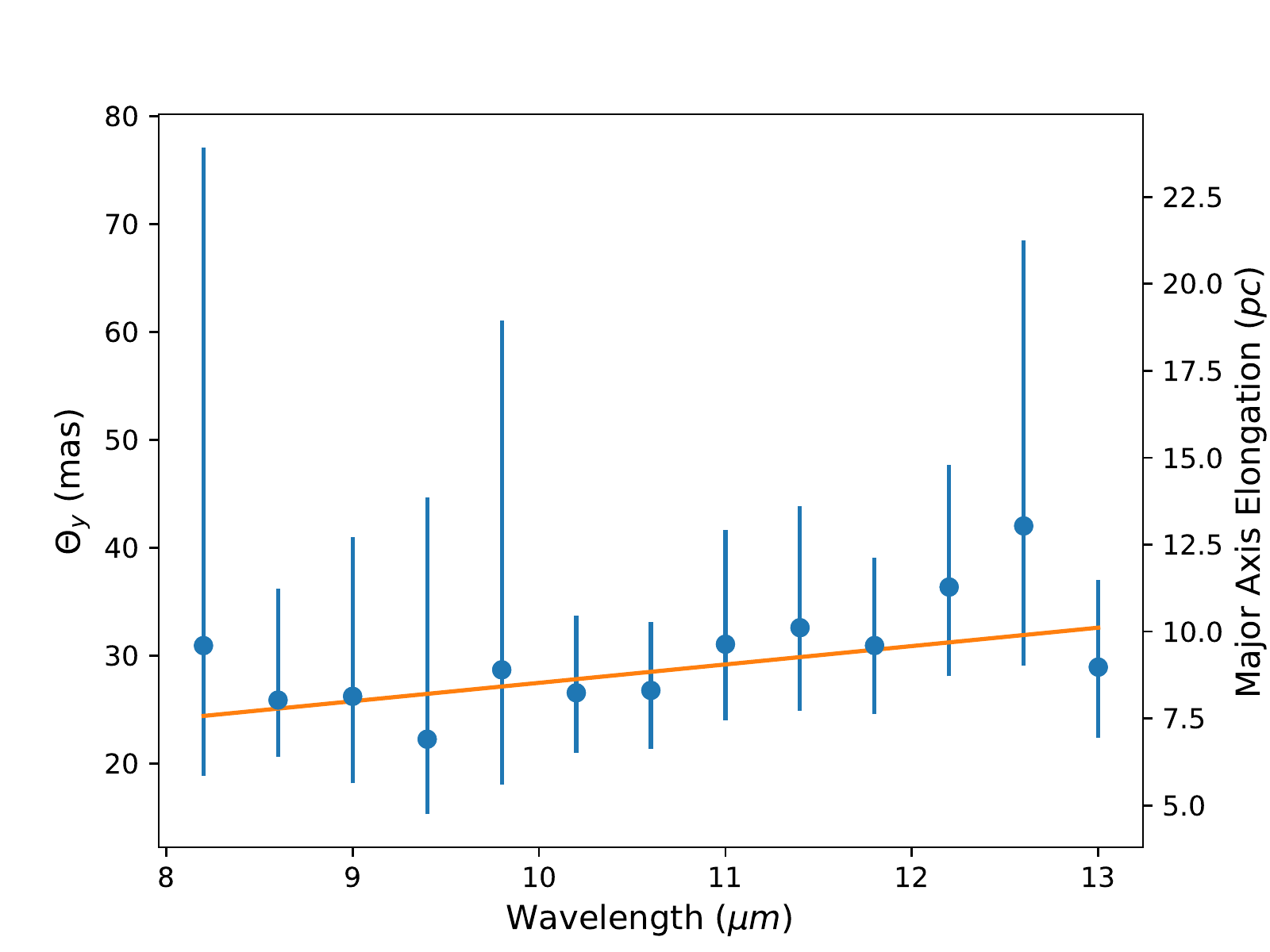}
\caption{Major axis FWHM, $\Theta_y$, of the geometric model for each wavelength bin. The orange line is the weighted line of best fit.}
\label{sigy-Lam}
\end{center}
\end{figure}

\begin{figure}[tbp!]
\begin{center}
\includegraphics[width=0.5\textwidth]{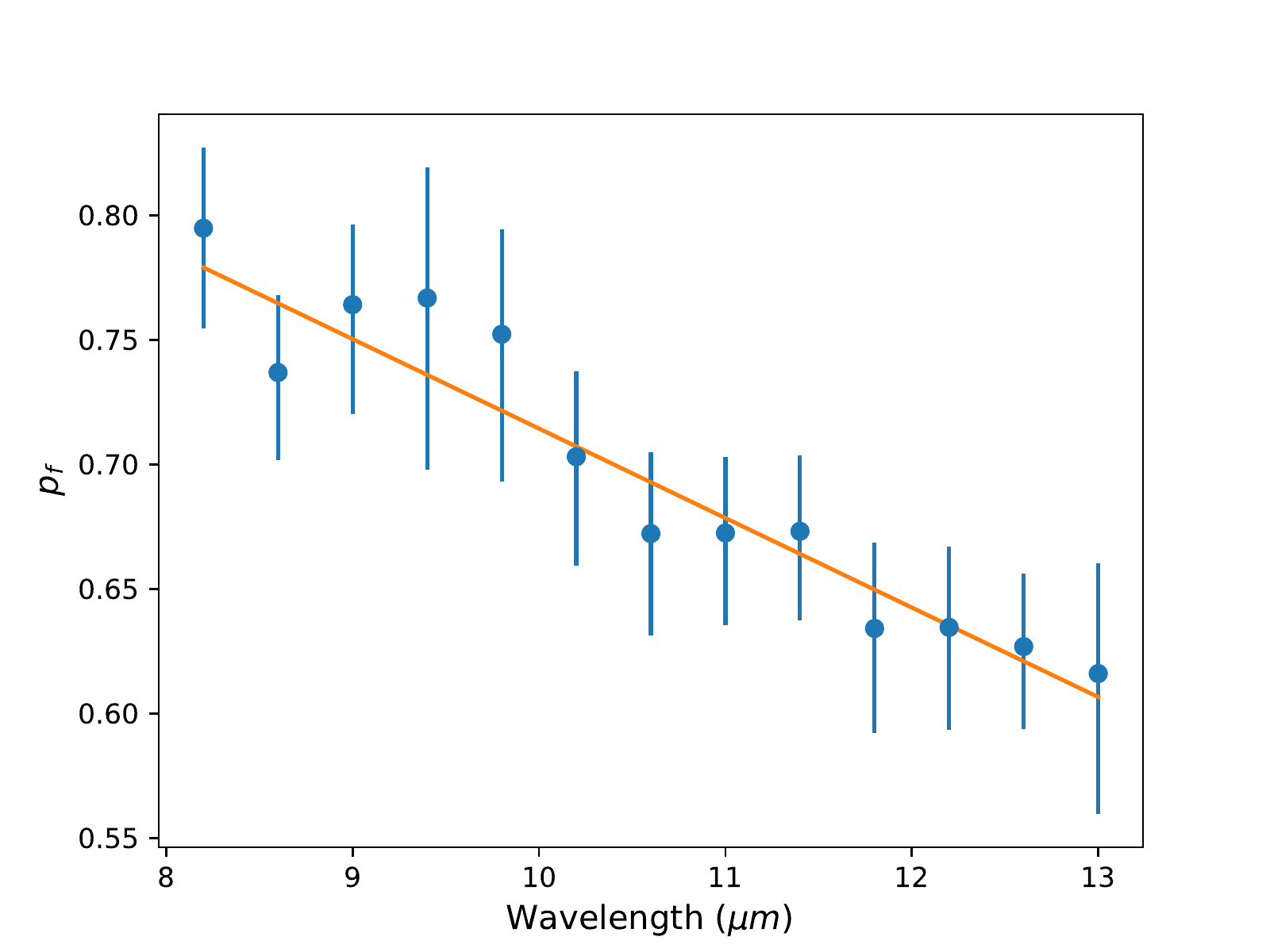}
\caption{Unresolved source fraction, $p_f$, of the geometric model for each wavelength bin. The orange line is the weighted line of best fit.}
\label{a-Lam}
\end{center}
\end{figure}

\begin{figure}[tbp!]
\begin{center}
\includegraphics[width=0.5\textwidth]{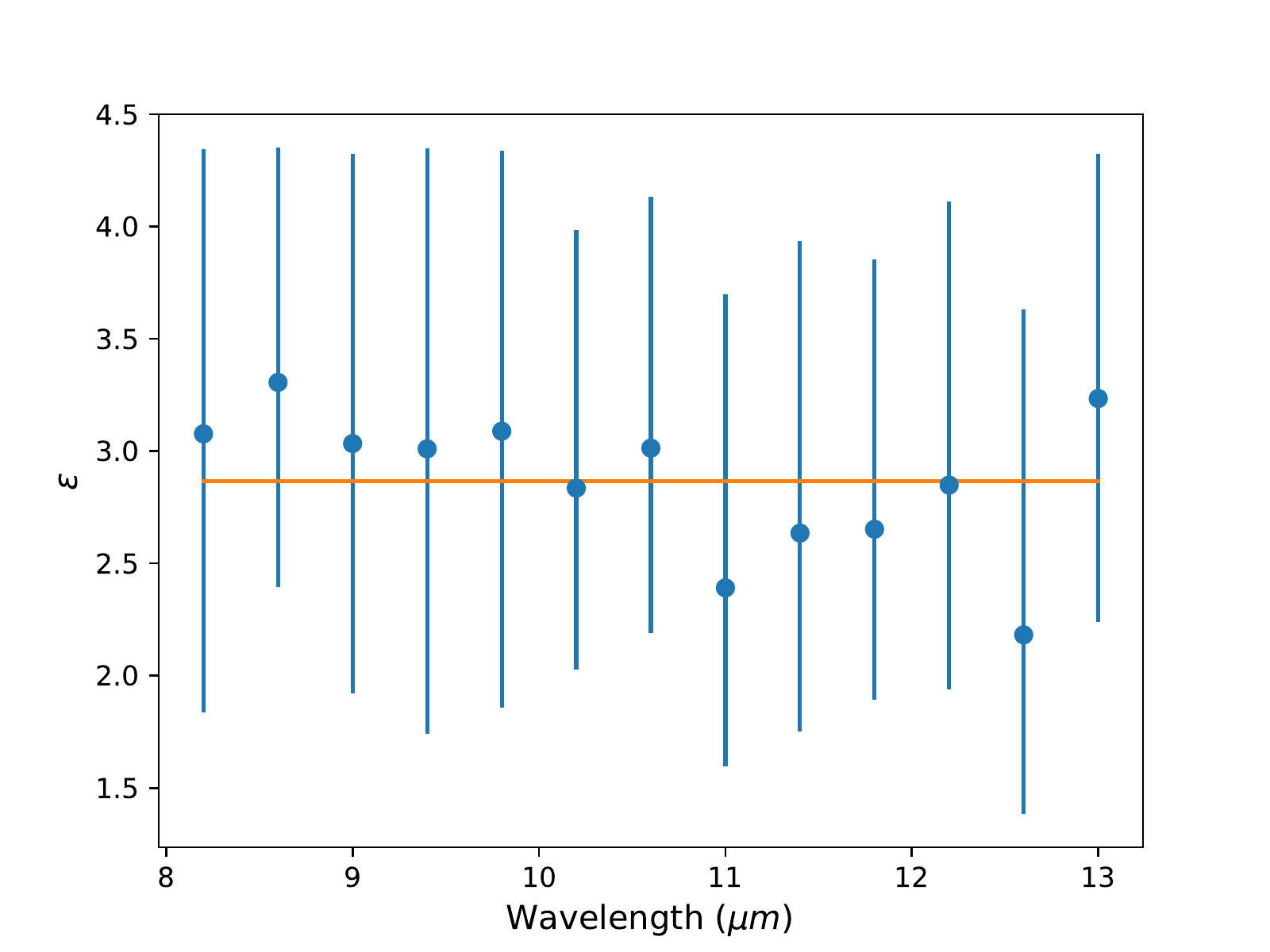}
\caption{Axis ratio, $\epsilon$, of the geometric model for each wavelength bin. The orange line is the weighted average.}
\label{r-Lam}
\end{center}
\end{figure}

We plot the model parameters as a function of wavelength in Figures \ref{PA-Lam}$-$\ref{r-Lam}, as well as the weighted average or a linear fit to the wavelength dependency. The weights are $1/\sigma_\mathrm{err}^2$.

In Figure \ref{PA-Lam}, we can see that the PA is fairly well constrained. We find that the PA has an average value of $155^\circ\pm14^\circ$. Figure \ref{PA-Lam} shows hints of a steady decrease in PA with increasing wavelength; however, there is no statistically significant correlation found, using the Spearman rank, with a p value of $0.45\pm0.3$. At the same time, in Figure \ref{a-Lam} we find that the unresolved source fraction increases from 62\% to 79\% for shorter wavelengths. We interpret this as greater influence from the hot component at shorter wavelengths. We also find that the major axis is well defined above $10\,\mu\mathrm{m}$ with $\Theta_y=26.6\,$mas at $10.2\,\mu\mathrm{m}$. The major axis increases with wavelength to $\Theta_y=32.4\,$mas at $13.0\,\mu\mathrm{m}$ (Figure \ref{sigy-Lam}). Because the longer wavelengths probe cooler dust, we see this as evidence for the dust being heated by the central engine. We find $\epsilon(\lambda)=2.9\pm0.3$, without any dependency on wavelength. This suggests that the geometry does not change significantly with wavelength, only the total size changes. To highlight any change in geometry with wavelength, we plot an RGB image of our geometric model results in Figure \ref{66lightrad} together with the 66\% light radius. Each channel in this image represents the average of two of the model results: the blue represents the average of the $8.6\,\mu$m and $9\,\mu$m results; the green represents the average of the $10.2\,\mu$m and $10.6\,\mu$m results; and the red represents the average of the $11.8\,\mu$m and $12.2\,\mu$m results. Each channel is convolved with the beam size at that wavelength. The two component structure with the strongly elongated extended component can be clearly seen. The above mentioned increase in size of the extended component with wavelength can also be seen: the emission starts bluer in the centre and becomes redder toward the edges of the extended component.

Figure \ref{Vis-BL mod} shows the interferometric data at $11.8\,\mu$m as a function of BL, similar to Figure \ref{fig BL ext}, with the geometric model overplotted. This plot suggests that the PA is not constrained solely by the single visibility measurement at $\mathrm{BL}=56.3$\,m and $\mathrm{PA}=353\fdg2$ which has a lower visibility than other points at similar baseline lengths (red point in Figure 3). To further check this, we re-fitted the $11.8\,\mu\mathrm{m}$ data without this observation. We recovered the PA as ${158^{+17}_{-9}}^\circ$, the unresolved fraction as $0.63^{+0.03}_{-0.04}$, $\epsilon$ as $2.2^{+1.2}_{-0.7}$, and a $\Theta_y$ of $27.8^{+8.6}_{-6.5}\,$mas. The result is within errors of the original fit but with greater uncertainty. From this, we can say the point helped to constrain the model result but is not solely responsible for it. The most sensitive result of the change is the PA. Each observation at the \textasciitilde$55$ meter baseline length constrains the direction of the extension. The errors in Figure \ref{PA-Lam} only include the uncertainties found by MCMC fitting and do not include the uncertainty found from the influence of the \textit{uv} plane.

\begin{table*}
\caption{Results of model fitting for each wavelength bin.}
\begin{tabular*}{\textwidth}{c | @{\extracolsep{\fill}} c c c c c}\hline

Wavelength ($\mu \mathrm{m}$)&$p_f$&$\epsilon$&$\Theta_y (\mathrm{mas})$&$\theta$ ($\deg$)&$\ln(f)$\\ \hline \hline
8.2&$0.79^{0.03}_{0.04}$&$3.08^{1.27}_{1.24}$&$30.9^{46.2}_{12}$&$170^{30}_{47}$&$-8.9^{4.3}_{4.1}$\\
8.6&$0.74^{0.03}_{0.04}$&$3.31^{1.05}_{0.91}$&$25.9^{10.3}_{5.3}$&$160^{19}_{13}$&$-9.8^{4.6}_{3.7}$\\
9&$0.76^{0.03}_{0.04}$&$3.03^{1.29}_{1.11}$&$26.2^{14.8}_{8.1}$&$150^{18}_{14}$&$-9^{3.9}_{4.1}$\\
9.4&$0.77^{0.05}_{0.07}$&$3.01^{1.34}_{1.27}$&$22.3^{22.4}_{6.9}$&$182^{30}_{27}$&$-10.8^{5}_{3.1}$\\
9.8&$0.75^{0.04}_{0.06}$&$3.09^{1.25}_{1.23}$&$28.7^{32.4}_{10.6}$&$155^{32}_{33}$&$-9.3^{4.3}_{3.9}$\\
10.2&$0.7^{0.03}_{0.04}$&$2.83^{1.15}_{0.81}$&$26.6^{7.1}_{5.6}$&$155^{13}_{8}$&$-10.5^{4.7}_{3.4}$\\
10.6&$0.67^{0.03}_{0.04}$&$3.01^{1.12}_{0.82}$&$26.8^{6.3}_{5.4}$&$160^{14}_{7}$&$-10.8^{4.6}_{3.2}$\\
11&$0.67^{0.03}_{0.04}$&$2.39^{1.31}_{0.8}$&$31.1^{10.6}_{7}$&$152^{12}_{12}$&$-9.3^{4}_{3.9}$\\
11.4&$0.67^{0.03}_{0.04}$&$2.63^{1.3}_{0.88}$&$32.6^{11.3}_{7.7}$&$154^{14}_{11}$&$-9.6^{4.2}_{3.7}$\\
11.8&$0.63^{0.03}_{0.04}$&$2.65^{1.2}_{0.76}$&$30.9^{8.1}_{6.4}$&$155^{16}_{7}$&$-10.6^{4.6}_{3.3}$\\
12.2&$0.63^{0.03}_{0.04}$&$2.85^{1.26}_{0.91}$&$36.4^{11.3}_{8.3}$&$149^{8}_{8}$&$-9.2^{4}_{4}$\\
12.6&$0.63^{0.03}_{0.03}$&$2.18^{1.45}_{0.8}$&$42^{26.5}_{12.9}$&$140^{22}_{21}$&$-9.6^{4.3}_{3.7}$\\
13&$0.62^{0.04}_{0.06}$&$3.23^{1.09}_{0.99}$&$28.9^{8.1}_{6.5}$&$170^{19}_{13}$&$-10^{4.6}_{3.6}$\\

 \hline
\end{tabular*}
\tablecomments{$p_f$ is the unresolved source fraction, $\epsilon$ is the ratio of the major axis to minor axis, $\Theta_y$ is the major axis FWHM, $\theta$ is the angle from north to east of the $\Theta_y$ component of the Gaussian and $f$ is the fractional amount for which the variance is underestimated by the likelihood function if the errors were assumed correct \citep{foreman-mackey_emcee:_2013}.}
\label{Tresult}
\end{table*}

\begin{table}
\caption{Results of the best fit \textit{CAT3D-WIND} model}
\begin{tabular*}{0.47\textwidth}{c | @{\extracolsep{\fill}} c c c c c c c c}\hline
Parameter&$N_0$&$a_d$&$a_w$&$h$&$\theta_w$&$\theta_\sigma$&$f_{\mathrm{wd}}$&inc\\ \hline \hline
Value&$5$&$-3$&$-1$&$0.1$&$30\degr$&$10\degr$&$0.6$&$60\degr$\\ \hline
\end{tabular*}
\label{T Rad mod}
\tablecomments{$N_0$ is the average number of clouds in the line of sight in the equatorial region, $a_d$ is the radial power-law index for the disk $\alpha r^{a_d}$, where $r$ is in units of the sublimation radius, $a_w$ is the radial power-law index of the dust clouds in the polar wind, $h$ is a unitless disk height scaling factor, $\theta_w$ is the opening angle of the polar wind, $\theta_\sigma$ is the angular width of the polar wind, $f_{\mathrm{wd}}$ is the wind to disk ratio, and inc is the inclination.}
\end{table}

\subsection{CAT3D-WIND Radiative Transfer Modelling}

With more and more interferometric sources being discovered that show polar-elongated mid-IR emission, a new radiative transfer model of the parsec-scale dusty environment has recently been developed. \citet{honig_dusty_2017} present their \textit{CAT3D-WIND} model that consists of a dusty disk and a hollow-cone outflow. The parametrisation is a minimal extension to a more generic clumpy torus model and successfully reproduced the mid-IR interferometry and IR SED of NGC 3783, simultaneously.

\begin{figure}
\includegraphics[width=0.4\textwidth, trim={1cm 0 2cm 0}]{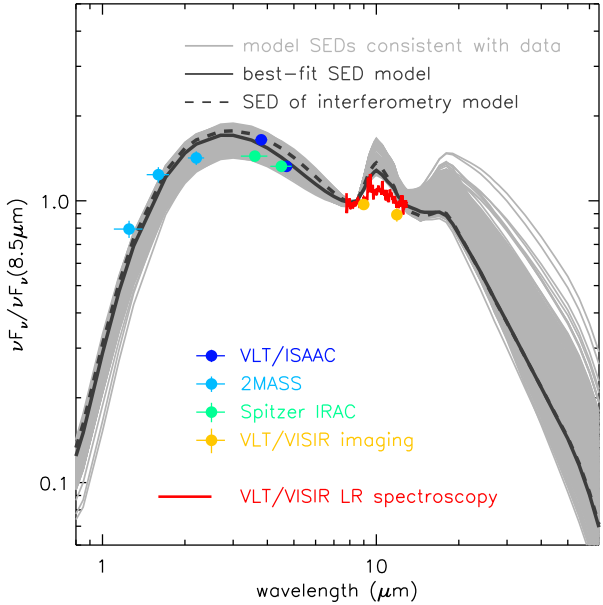}
\caption{High-resolution SED of ESO\,323-G77. The filled circles represent observed photometry from the near-IR and mid-IR, while the red line shows the VISIR spectrum. The data have been corrected for Galactic extinction of $A_V=1.2$. The light-grey lines show \textit{CAT3D-WIND} model SEDs that are consistent with the 95\% confidence interval of the best fit (solid dark-grey line; see text for details). The dashed black line shows the SED of the model that is used to reproduce the MIDI interferometry in Figure~\ref{cat3dvis}.}
\label{sed}
\end{figure}

We followed the same strategy to model the interferometry of ESO\,323-G77 as \citet{honig_dusty_2017}. First, we use the high-resolution IR photometry and spectroscopy of ESO\,323-G77 to find acceptable fits to the SED from the suite of approximately 132,000 \textit{CAT3D-WIND} models. Figure \ref{sed} shows the observed IR SED in comparison to model SEDs. VLT/ISAAC fluxes have been extracted from archival ESO data\footnote{data acquired as part of programme 290.B-5113, PI D. Asmus} using standard photometric procedures. The VISIR fluxes were presented by \citet{asmus_subarcsecond_2014} while the \textit{Spitzer} IRAC and 2MASS photometry was reported by \citet{kishimoto_mapping_2011}. We also adopt the latter paper's Galactic extinction correction of $A_V=1.2\,\mathrm{mag}$. The light-grey lines in Figure \ref{sed} represent all models within the 95\% confidence interval of the best SED fit (solid dark-grey line) according to a $\chi^2$ distribution. The SEDs strongly prefer compact disk emission ($a_d = -2.5\cdots-3$) with a more extended wind component ($-0.5 \le a_w \le -1.5$). The scale height, $h$, of the disk is relatively small, with three-quarter of all models preferring $h=0.1$ and the remainder $h=0.2$. Based on the SED alone, the mean number of clouds, $N_0$, along the line of sight is favoured as being $N_0=5$, with only half as many obscuring the AGN when grazing the hollow cone, i.e. $0.25 < f_\mathrm{wd} < 0.75$. Inclinations of these models range from $0^\circ$ to $60^\circ$ with a preference for lower inclinations. Finally, the opening angle of the cone is $\theta_w = 30^\circ$ in about 80\% of all acceptable models with the remaining at $\theta_w = 45^\circ$. The width of the hollow cone is unconstrained by the SEDs.

\begin{figure}
\begin{center}
\includegraphics[trim={1cm 0 0 0} ,clip,width=0.5\textwidth]{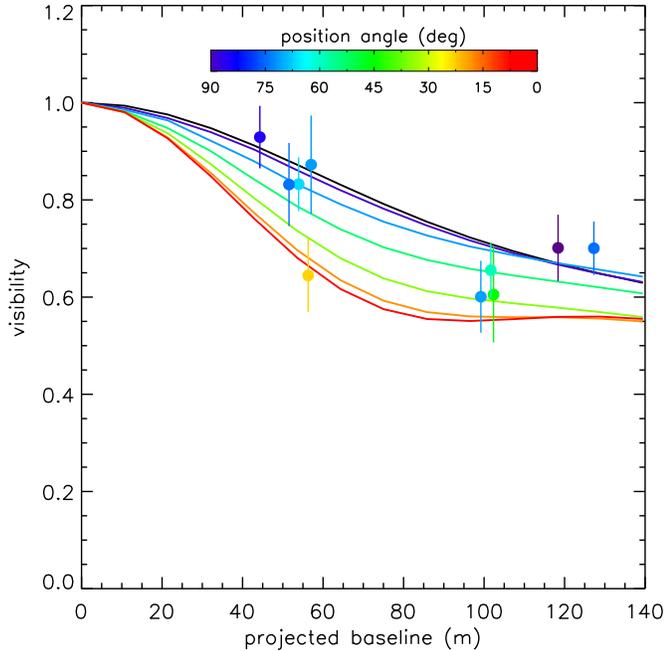}

\end{center}
\caption{Radial visibility versus baseline as observed (filled circles) and modelled with \textit{CAT3D-WIND} (solid lines) model at $12\,\mu$m for ESO\,323-G77. The colours represent position angle with respect to $155^\circ$, the major axis of the geometric model.}
\label{cat3dvis}
\end{figure}

Simulating 3D radiative transfer model images and extracting visibilities is computationally expensive. As it turns out, one can easily exclude families of model parameter combinations as good fits to the interferometry data. Despite its nature as a type 1 AGN, ESO\,323-G77 has strongly elongated polar mid-IR emission. Low-inclination models are not able to reproduce this feature in visibilities since they do not show the required strong dependence of visibility on PA. For the same reason, opening angles of $45^\circ$ in the remaining models are disfavoured. Another set of models with extended radial dust distributions of $a_w = -0.5$ generally overestimate the contribution from the wind leading to visibilities that are too low. We then simulated model images in the remaining parameter space. Reproducing the correct level of unresolved emission (visibility $\sim$0.6) and the strong PA dependence leaves a very small parameter space. A satisfactory representation of visibilities has been achieved for the model parameters listed in Table \ref{T Rad mod}. The corresponding model visibilities at $12\,\mu$m are compared to the data in Figure \ref{cat3dvis}. Within the tested range of parameter steps, those parameters were the only ones to simultaneously reproduce the SED, overall visibility levels, and PA dependence. The only major degeneracy remained for the width of the hollow cone, $\theta_\sigma$, where values of $10^\circ$ and $15^\circ$ led to similar results.

Similarly to the \textit{CAT3D-WIND} modelling of NGC 3783 in \citet{honig_dusty_2017}, these new results illustrate the constraining power of interferometric data on the geometry of the dust distribution. SED modelling is restricted to the integrated, zero-dimensional emission, with the geometric distribution contributing a highly degenerate flux weighting function. Interferometry provides additional two-dimensional constraints of the flux distribution that reveal features not necessarily captured by a model based only on an SED. Indeed, the SED of ESO\,323-G77 has a shape that is covered by classical torus-only models as well as the disk+wind model \citep[see][for a comparison of torus model and disk+wind model SED parameter space]{honig_dusty_2017}. However, the clumpy torus model alone fails to simultaneously fit the SED and the interferometric data. Interferometry revealed the significant polar-elongated mid-IR emission, which can only be reproduced by the disk+wind model.

\section{Discussion}\label{Discussion}

\subsection{Comparison of the Results}

\begin{figure}[tb!]
\begin{center}
\includegraphics[scale=0.5]{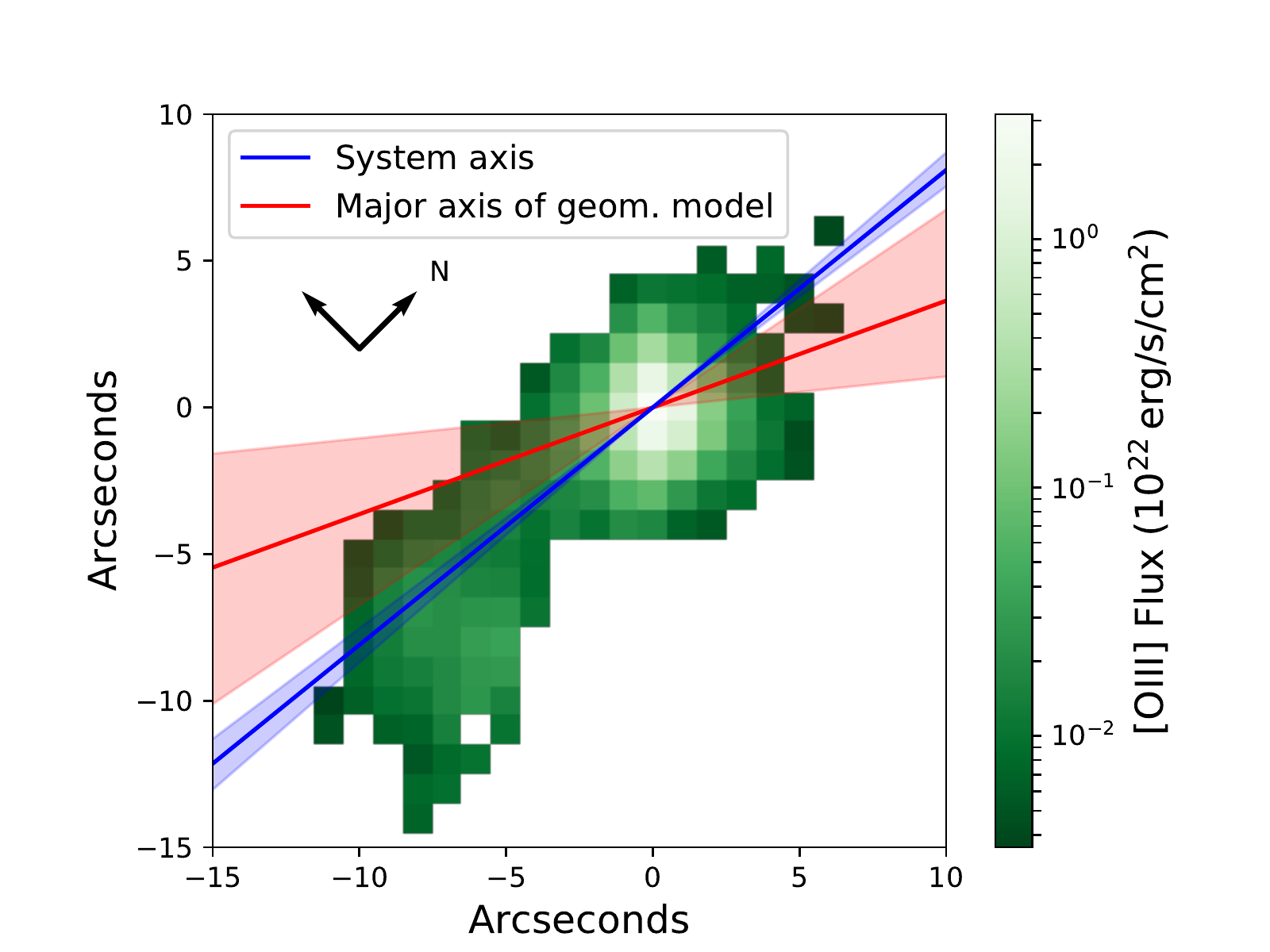}
\end{center}
\caption{Image of the 500.7\,nm [\ion{O}{3}] emission line for ESO\,323-G77 using data from S7 \citep{thomas_probing_2017}. North and East are given in the plot. The red line is the mean value found for the elongation direction of the geometric model in this paper and the blue line is the polar axis implied by the polarisation direction \citep{schmid_spectropolarimetry_2003}. The shaded regions are the 1$\sigma$ errors on the measurements. The axes are distance from the centre of the AGN.}
\label{O3}
\end{figure}

The 500.7\,nm [\ion{O}{3}] emission from the S7 data is shown in Figure \ref{O3}. We overplotted the PA of the geometric model as a red line and the polar axis derived from the polarisation measurements as a blue line. Although the NLR appears to be bent or obscured $5"$ from the central region, Figure \ref{O3} shows that the mid-IR emission in ESO\,323-G77 is polar extended.

The constraints on the mid-IR size presented here are consistent with previous observations of ESO\,323-G77: \citet{kishimoto_mapping_2011} report a Gaussian FWHM of $24.6\,$mas $\pm\ 0.8\,$mas, while we find an average FWHM of 21\,mas $\pm\ 2.3\,$mas at $11.8\,\mu\mathrm{m}$. \citet{burtscher_diversity_2013} perform a two component fit, consisting of a Gaussian and a point source, finding that their point source fraction for ESO\,323-G77 is unconstrained. They proceeded to use a 1 component model, consisting of only the radially symmetric Gaussian, and by consequence find an FWHM of only $6.74\,$mas. This is far smaller than our result and the result from \citet{kishimoto_mapping_2011}. \citet{lopez-gonzaga_mid-infrared_2016} report a FWHM of $17.3\,$mas for the minor axis and an $\epsilon$ of 1.4 in their Figure A.1 (N. Lopez 2017, private communication). This corresponds to an average size of 21\,mas, which agrees with our result within 1$\sigma$.

\begin{figure}[tbp]
\includegraphics[width=0.5\textwidth]{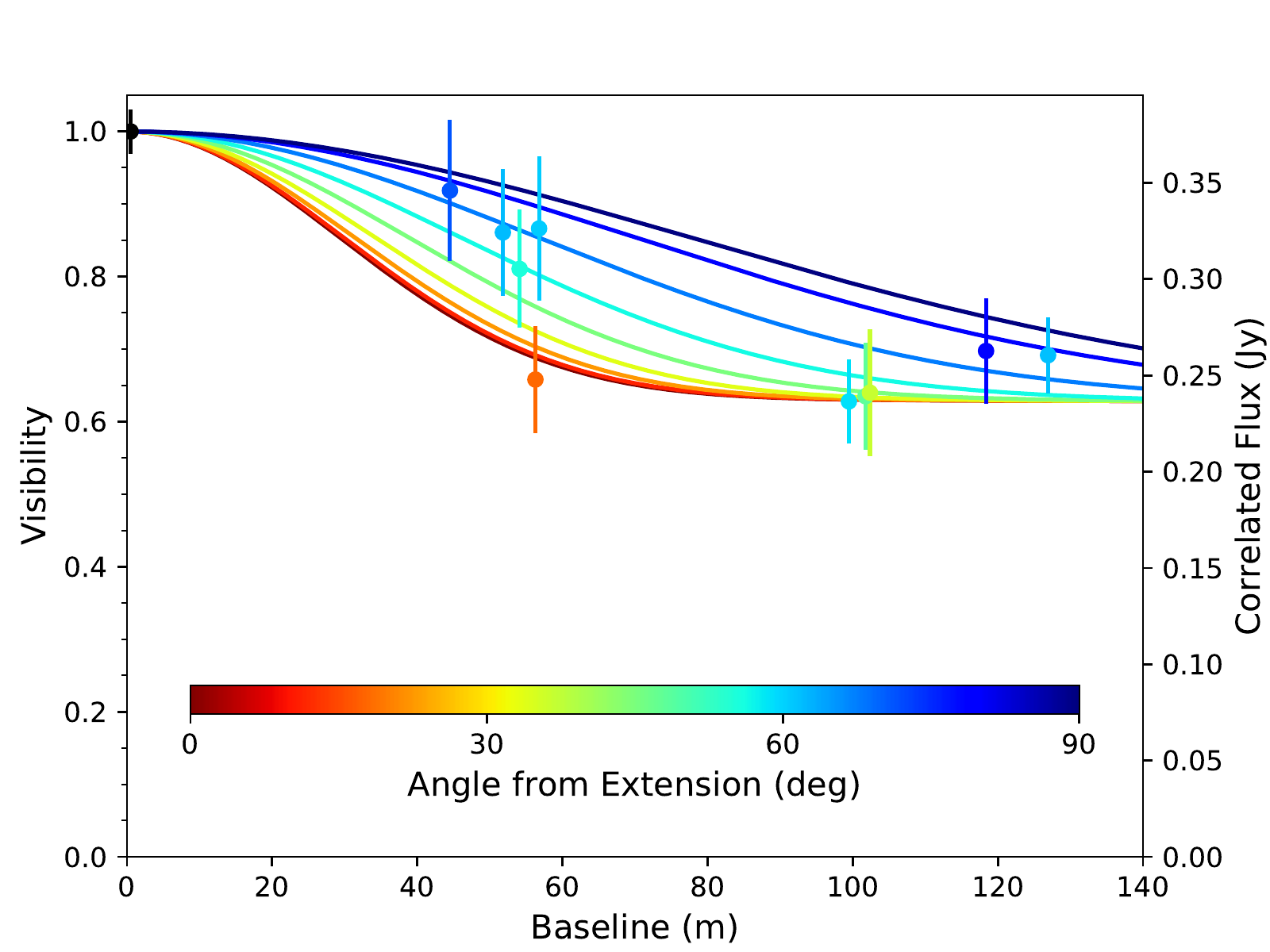}
\caption{The $V(11.8\,\mu\mathrm{m})$ of each observation against its baseline. The data points are coloured by PA from the Gaussian major axis of the geometric model in real space. The fitted geometric model is overplotted for different PAs. The value at 0\,m base length is the VISIR observation at $11.8\pm0.2 \,\mu\mathrm{m}$ and its errors are a systematic offset for all the \textit{uv} points.}
\label{Vis-BL mod}
\end{figure}

The geometric model reproduces the interferometric data very well (see Figure \ref{Vis-BL mod}), and directly infers the geometry of the source from the data. The \textit{CAT3D-WIND} model on the other hand gives a better description of the properties of the AGN beyond geometry as described in \citet{honig_dusty_2017}. The parameters between both the geometric and the radiative transfer models are very comparable. The primary feature constrained by the \textit{CAT3D-WIND} model is the average distribution and density of the dust clumps, which is much more concentrated toward the disk than for other AGN \citep[see, e.g.][]{honig_dusty_2017}. This is essential in understanding the luminosity distribution to make predictions for what we would observe in other wavelengths. Since this was generated using the SED, it did not relate to the interferometric data directly, instead the observed interferometric data narrowed down the possible \textit{CAT3D-WIND} models that fit the SED. Future observations in other wavelengths could further narrow down the possible models.

We can use the \textit{CAT3D-WIND} model to predict the size and shape of the \emph{hot} dust emission. Based on our best fit model, we expect a hot dust disk with an inclination of $60^\circ$ and a major axis FWHM, if approximated by a Gaussian, of about 0.3\,pc ($\approx1\,\mathrm{mas}$). This can be observed with near-IR interferometry in the $K$-band, e.g.\ with VLTI/GRAVITY, and corresponds to a visibility of 0.8 on a 130\,m baseline (UT1/UT4) oriented along the system plane at PA=$65^\circ$. In the direction of the outflow (i.e.\ along the system axis), we would expect a higher visibility of 0.93 at the same baseline length. The implied radius of the $K$-band emission from this model of 0.15\,pc is a factor of 2.3 larger than the model sublimation radius of $\sim$0.065\,pc inferred independently from the SED fitting and the visibility scaling. This is expected since interferometry measures an average size of the the brightness distribution rather than the smallest emission radius \citep{kishimoto_innermost_2007,kishimoto_innermost_2011}. Reverberation mapping, on the other hand, is more sensitive to the smallest scales of the brightness distribution. Indeed, reverberation mapping of ESO\,323-G77 shows a lag between the optical and near-IR emission of about 75 days or 0.063\,pc, fully consistent with our inference from the presented model (B. Boulderstone et al. 2018, in preparation).

\citet{lopez-gonzaga_mid-infrared_2016} report seven objects with a sufficient \textit{uv} coverage to constrain an extension. Of these objects, six are type 2 Seyferts and one is a type 1 Seyfert (NGC3783), making ESO\,323-G77 only the second type 1 Seyfert with sufficient \textit{uv} coverage to constrain an extension. The detection of a second polar extended type 1 is in strong support of the disk+wind model because polar extensions in type 2s can also be explained by the standard torus model due to self-shielding effects or anisotropic illumination \citep{honig_parsec-scale_2012}; however, polar extended type 1 AGN cannot easily be explained by such models. 

In fact, \citet{lopez-gonzaga_mid-infrared_2016} find that the PA of all the objects that show an extension is approximately $20^\circ$ away from the system axis. We also find this for ESO\,323-G77. The disk+wind model explains this as an edge brightening effect of the hollow cone.

\subsection{An Evolution of the Dust Distribution with Eddington Ratio?}

Our observations and modelling have revealed highly elongated mid-IR dust emission in ESO\,323-G77 with a much higher elongation than those found in previous work \citep{honig_parsec-scale_2012,lopez-gonzaga_mid-infrared_2016}. Furthermore, ESO\,323-G77 is the first clear example of an AGN dominated by unresolved emission in the mid-infrared that, at the same time, has a detected polar extension.

From the full IR SED and the strong unresolved source-wavelength anti-correlation revealed by the geometric modelling, we interpret this dominant unresolved source as the Rayleigh-Jeans tail of the hot dust emission originating from the inner region of the disk close to the sublimation radius. These results indicate strong similarities of ESO\,323-G77 with the compact emission seen in two quasars \citep{kishimoto_mapping_2011}. Given that ESO\,323-G77 is considered to have a higher Eddington ratio than typical Seyferts (relatively narrow Hydrogen emission lines in the optical), similar to quasars, our result hints at evolution of the geometric dust distribution around an AGN with Eddington ratio. Indeed, such a scenario is consistent with radiation pressure shaping the mass distribution on the sub-parsec to parsec-scale environment of the AGN \citep{fabian_effect_2008,ricci_close_2017}.

\section{Summary} \label{Summary}

We presented a detailed study of the type 1.2 Seyfert AGN ESO\,323-G77 on the milliarcsecond scale in the mid-IR using MIDI and VISIR. We studied the geometry of the $300-400\,$K dust using a simple geometric model as well as the radiative transfer model \textit{CAT3D-WIND} \citep{honig_dusty_2017}. The system axis of ESO\,323-G77, from polarisation measurements and the NLR, is $174^\circ\pm2^\circ$ \citep{schmid_spectropolarimetry_2003}. 

\begin{enumerate}

\item The geometric modelling of the interferometric data revealed that $\gtrsim62\%$ of the flux is unresolved at all wavelengths and baselines. Among the remaining, resolved, fraction of the emission we discovered a dust structure with an axis ratio $\epsilon=2.9\pm0.3$ elongated along $\mathrm{PA}=155^\circ\pm14^\circ$. Comparing this dust extension to the system axis in Figure \ref{O3}, we conclude that this extension is in the polar direction.

\item We tested the dependence of the geometric model on the \textit{uv} plane and showed that we could recover the PA reliably with an uncertainty of $14^\circ$ for $\epsilon\geqslant2$.

\item We created a model of ESO\,323-G77 with the \textit{CAT3D-WIND} model using its SED. Comparing the successful models to the interferometric data, the number of viable models is greatly narrowed down and we are left with the model described in Table \ref{T Rad mod}.
\end{enumerate}
We conclude that the geometric and \textit{CAT3D-WIND} models agree within error and that both are a good description of the structure of the $300-400\,$K dust in ESO\,323-G77. With the polar extension discovered in this paper there are a total of eight AGN which have a sufficient \textit{uv} coverage to constrain any extension. Six of these are polar extended and none have equatorial extension, making a strong case for the disk+wind model of AGN. In future observations, we hope to further test the \textit{CAT3D-WIND} model in the near-IR.

\acknowledgements
We would like to thank the referee for the thorough comments that helped improved our paper. JHL, SFH, and DA acknowledge support from the Horizon 2020 ERC Starting Grant \textit{DUST-IN-THE-WIND} (ERC-2015-StG-677117). PG thanks STFC for support (grant reference ST/J003697/2).
MK acknowledges support from JSPS under grant 16H05731.
This work made use of the KUBEVIZ software which is publicly available at \url{http://www.mpe.mpg.de/~dwilman/kubeviz/}

\bibliographystyle{apj.bst}
\bibliography{Zotero.bib}

\appendix
\begin{figure}[b]
\includegraphics[height=17.5cm]{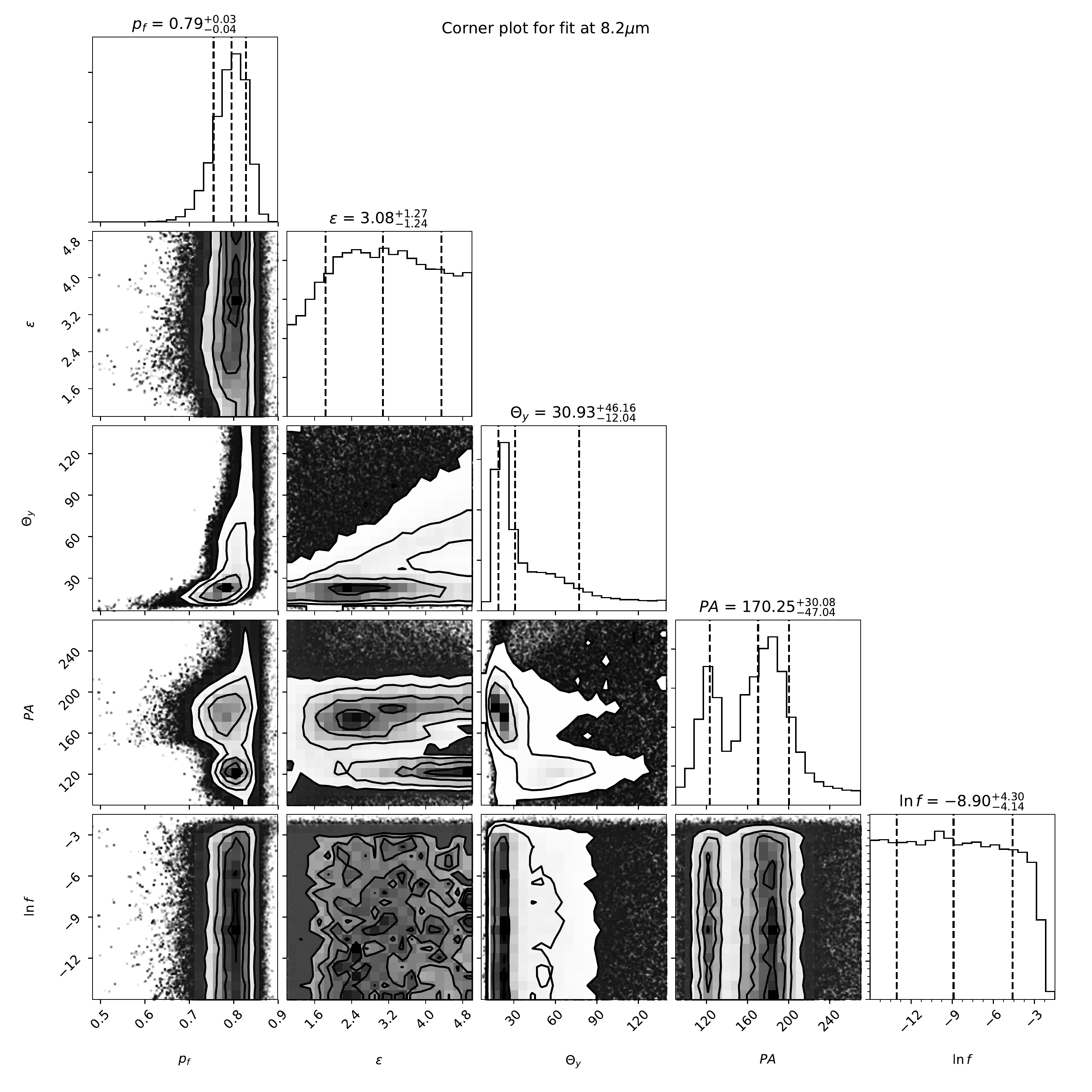}
\llap{\raisebox{12.1cm }{\includegraphics[height=4cm ,trim={2cm 4cm 2cm 6.2cm},clip]{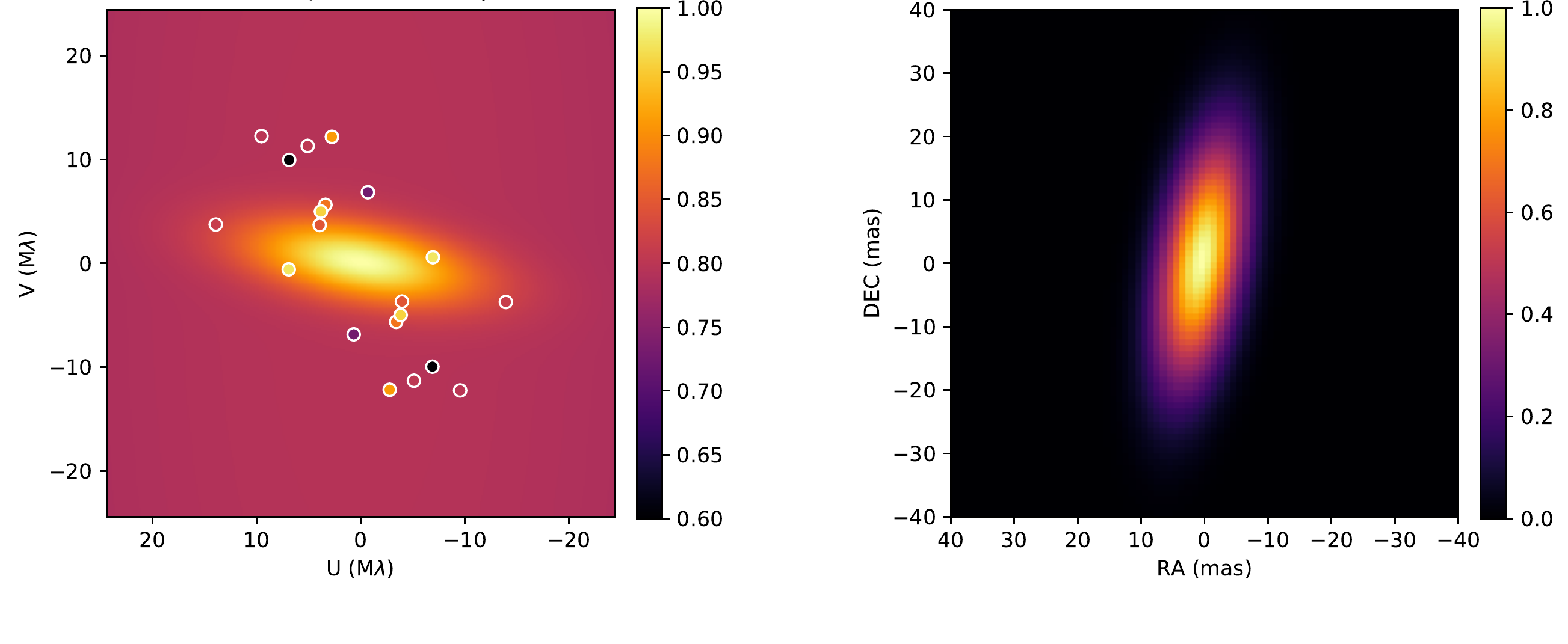}}}
\caption{Corner plot \citep{foreman-mackey_corner.py:_2016} of the Probability Density Function (PDF) of each parameter, the fitted Fourier space visibility distribution, and the reconstructed brightness distribution for the $8.2\,\mu$m geometric model fit.}
\label{8.2}
\end{figure}
\begin{figure}[b]
\includegraphics[height=17.5cm]{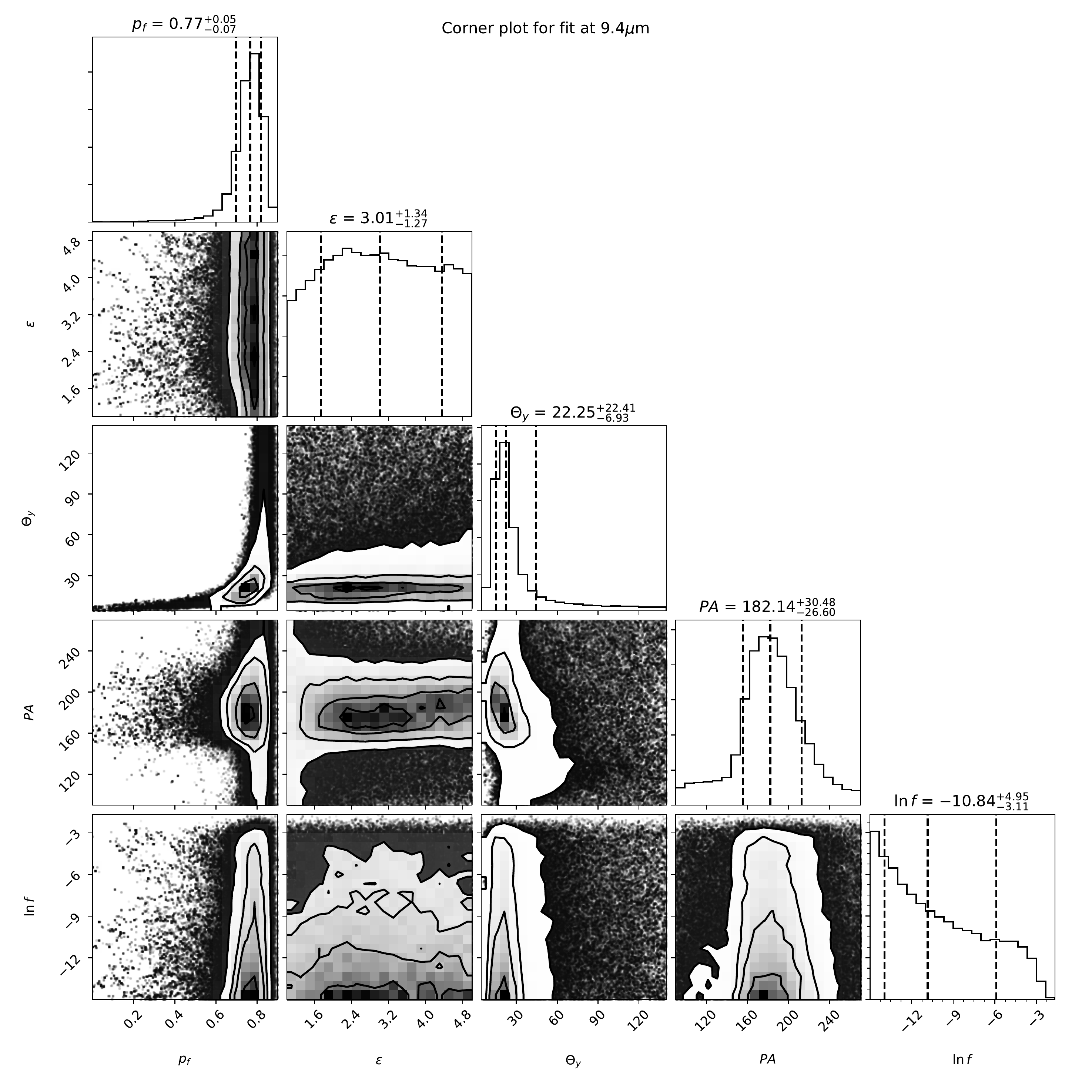}
\llap{\raisebox{12.1cm}{\includegraphics[height=4cm ,trim={2cm 4cm 2cm 6.2cm},clip]{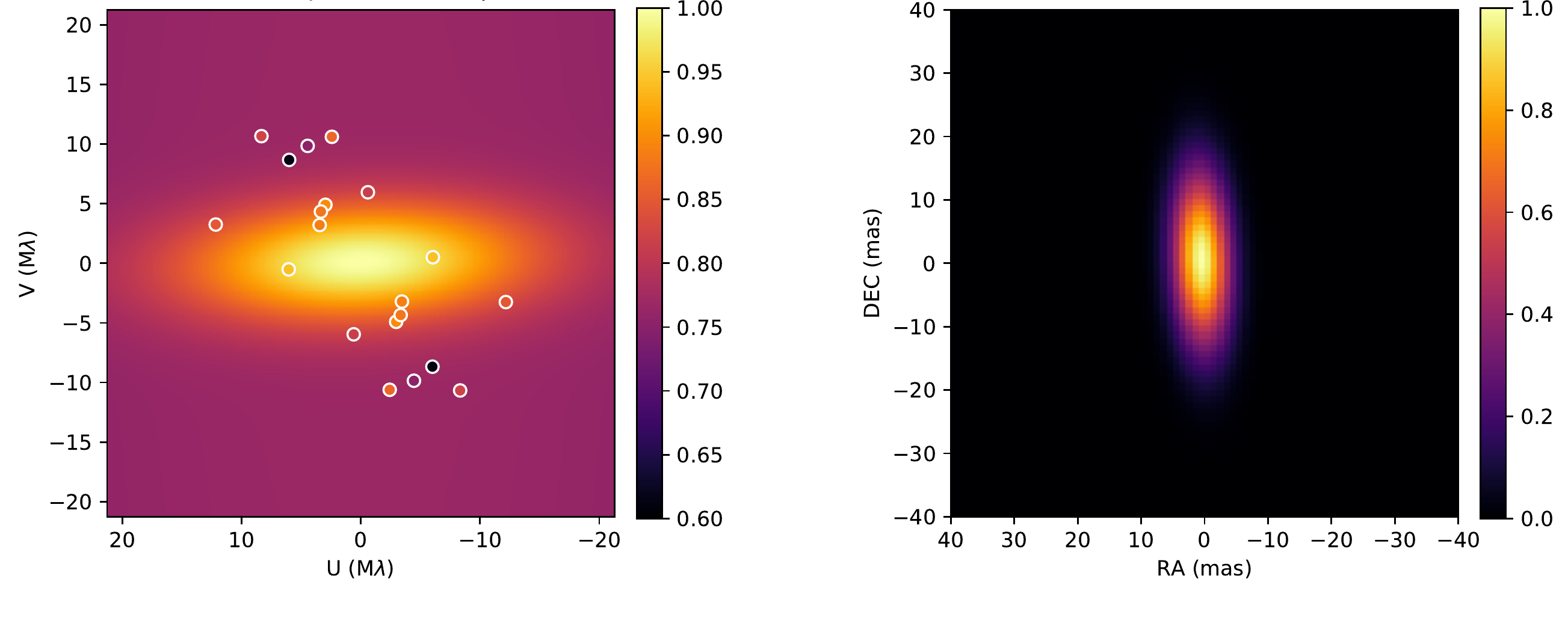}}}
\caption{Same as Figure \ref{8.2}, but for the $9.4\,\mu$m geometric model fit.}
\label{9.4}
\end{figure}
\begin{figure}[b]
\includegraphics[height=17.5cm]{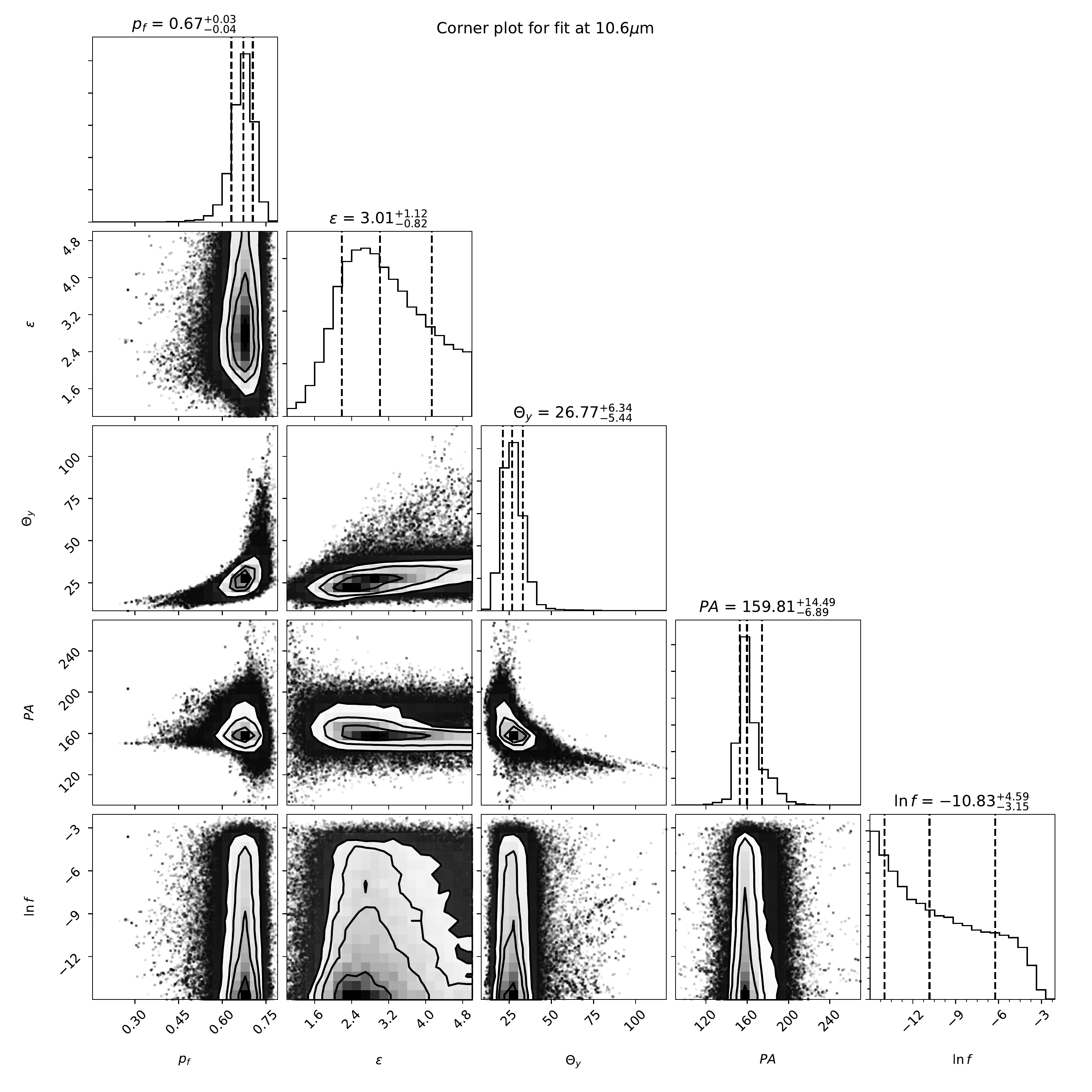}
\llap{\raisebox{12.1cm}{\includegraphics[height=4cm ,trim={2cm 4cm 2cm 6.2cm},clip]{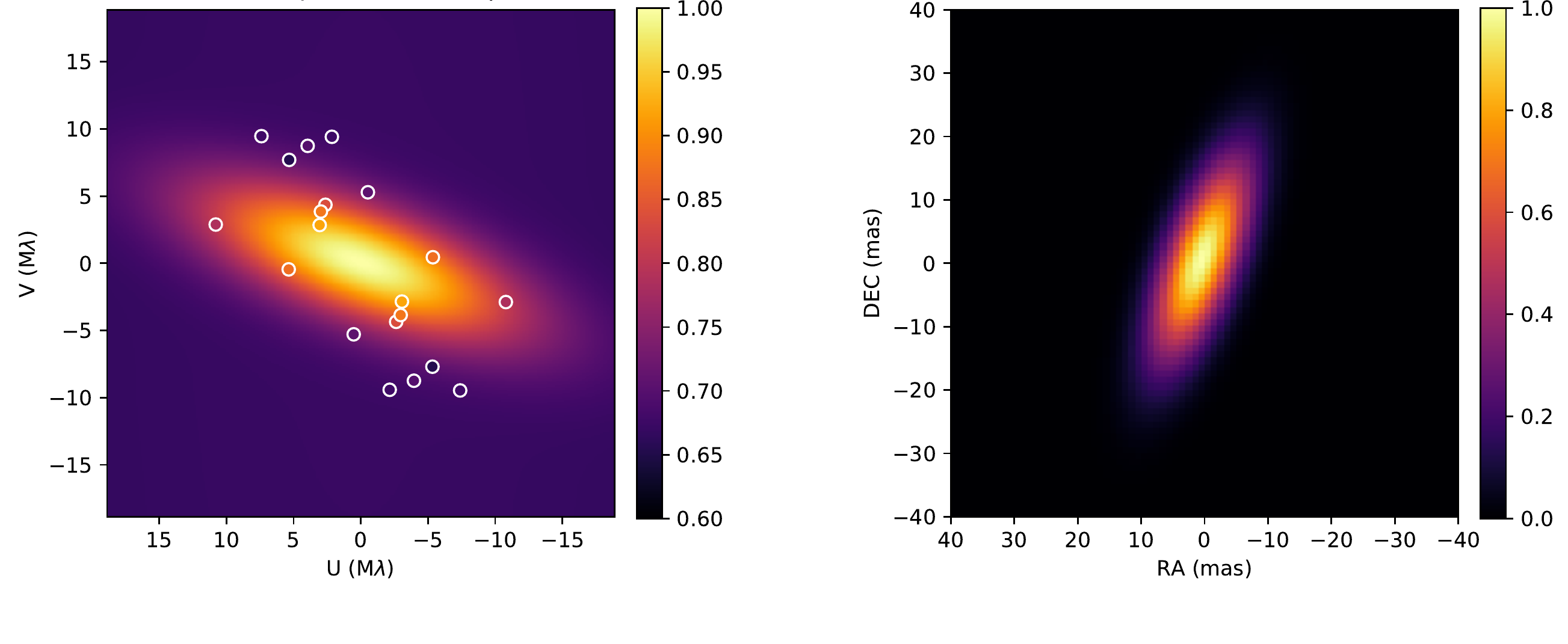}}}
\caption{Same as Figure \ref{8.2}, but for the $10.6\,\mu$m geometric model fit.}
\label{10.6}
\end{figure}
\begin{figure}[b]
\includegraphics[height=17.5cm]{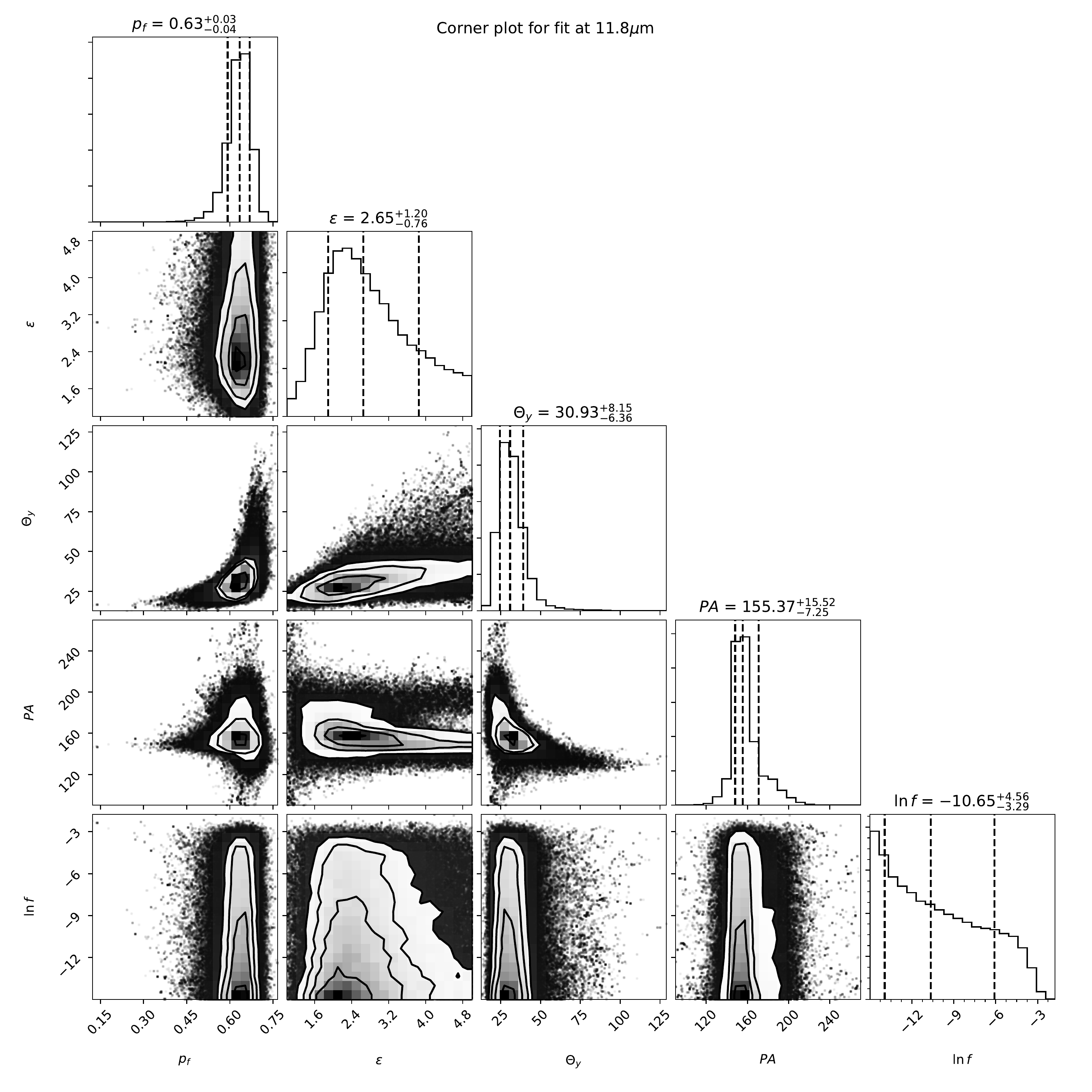}
\llap{\raisebox{12.1cm}{\includegraphics[height=4cm ,trim={2cm 4cm 2cm 6.2cm},clip]{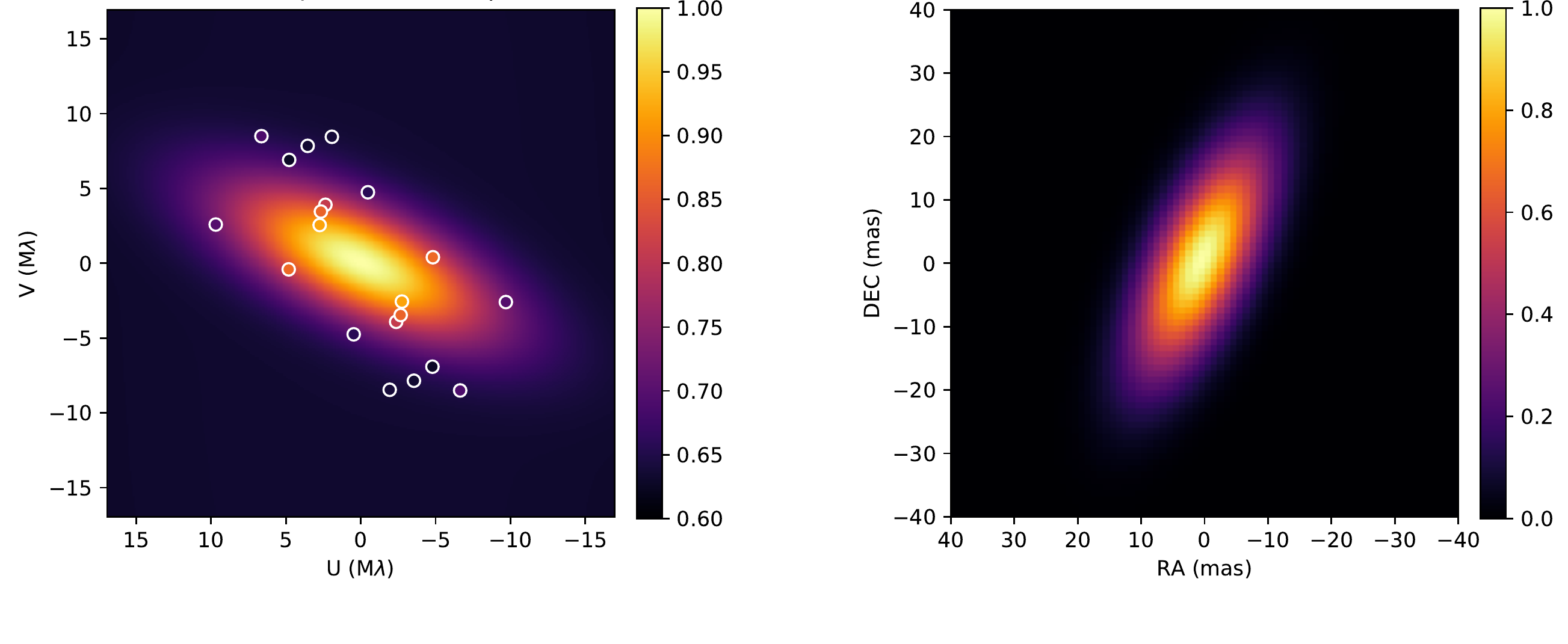}}}
\caption{Same as Figure \ref{8.2}, but for the $11.8\,\mu$m geometric model fit.}
\label{11.8}
\end{figure}
\begin{figure}[b]
\includegraphics[height=17.5cm]{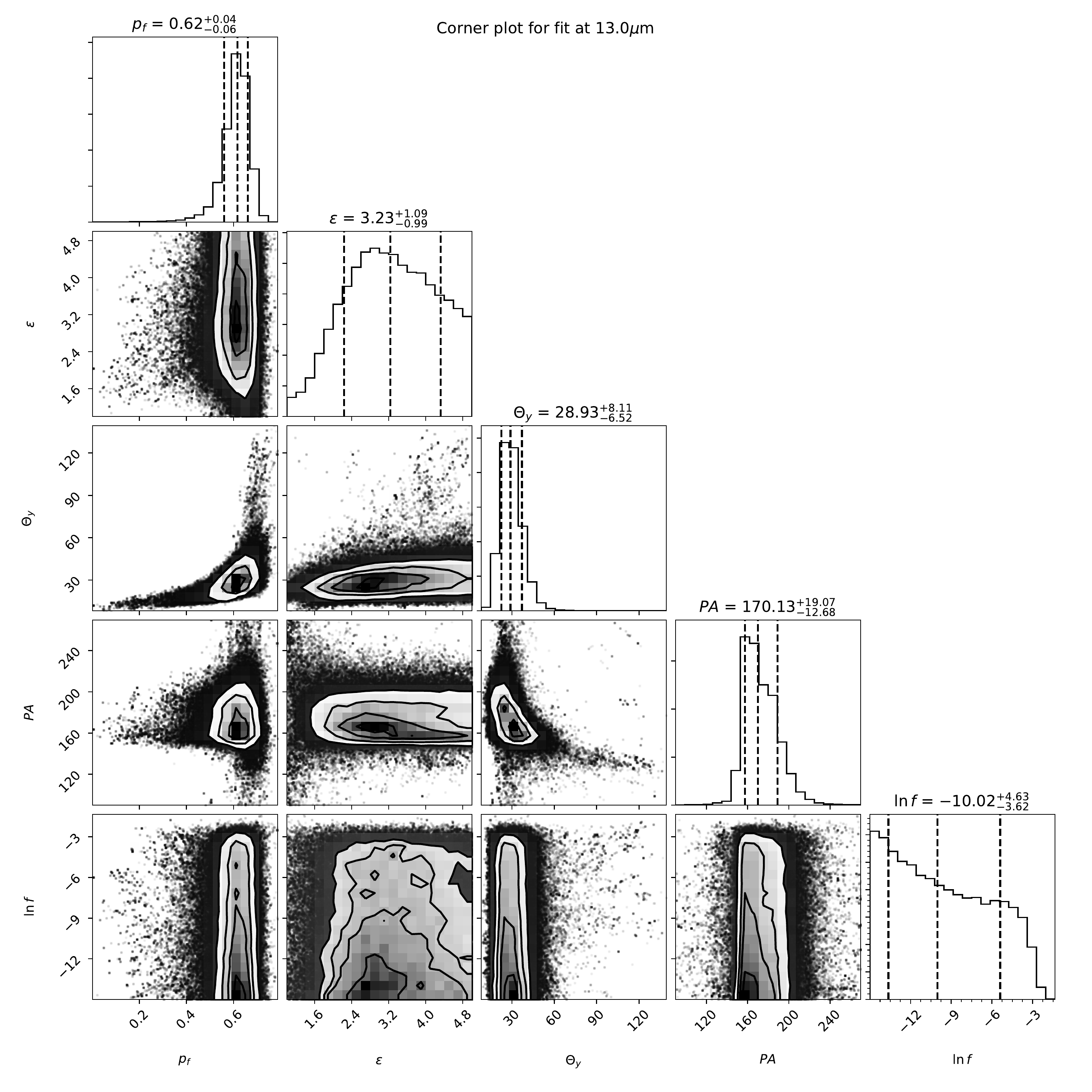}
\llap{\raisebox{12.1cm}{\includegraphics[height=4cm ,trim={2cm 4cm 2cm 6.2cm},clip]{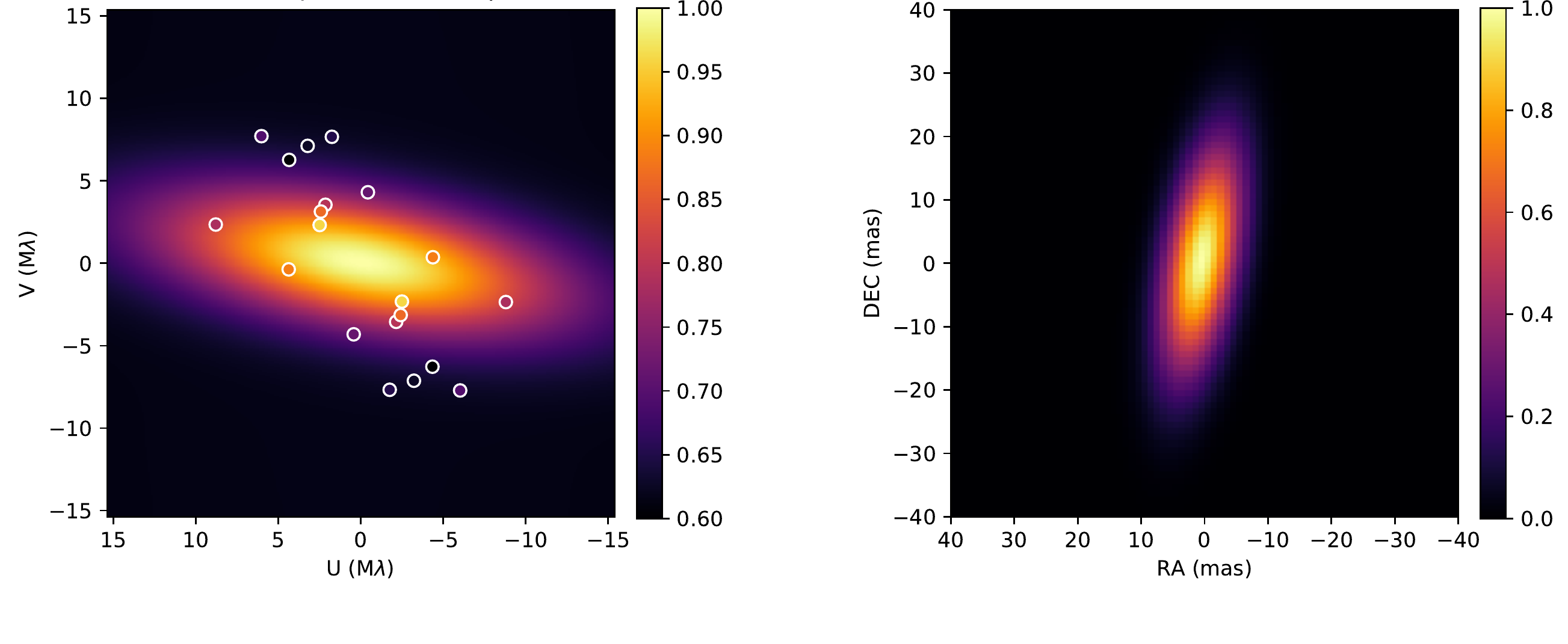}}}
\caption{Same as Figure \ref{8.2}, but for the $13\,\mu$m geometric model fit.}
\label{13.0}
\end{figure}
\end{document}